\documentclass[oldversion]{aa}
\usepackage{txfonts}
\usepackage{graphicx}
\usepackage{natbib}
\bibpunct{(}{)}{;}{a}{}{,}

\def\del#1{{\bf (DELETED TEXT)}}

\newcommand{\e}{{\rm e}}
\newcommand{\p}{{\rm p}}
\newcommand{\me}{m_\e}
\renewcommand{\mp}{m_\p}
\newcommand{\CR}{{\rm CR}}
\renewcommand{\th}{{\rm th}}

\newcommand{\dps}{\displaystyle}
\newcommand{\B}{{\mathcal B}}
\newcommand{\M}{{\mathcal M}}
\newcommand{\dd}{\mathrm{d}}
\newcommand{\cre}{\mathrm{CRe}}

\newcommand{\eps}{\varepsilon}
\newcommand{\vecbf}{\vec}
\newcommand{\vel}{\upsilon}
\newcommand{\bvel}{\vec{v}}

\newcommand{\upi}{\pi}
\newcommand{\umu}{\mu}
\def\rmn#1{{\rm #1}}

\title{Cosmic ray physics in calculations of cosmological
  structure formation}

\titlerunning{Cosmic ray physics in cosmological structure formation}

\author{T.~A.~En{\ss}lin \inst{1}, C.~Pfrommer \inst{1} \inst{2}, 
V.~Springel \inst{1}, \and M.~Jubelgas \inst{1}}
\offprints{T.~A.~En{\ss}lin}

\authorrunning{En{\ss}lin, et al.}

\institute{Max-Planck-Institut f\"{u}r Astrophysik,
  Karl-Schwarzschild-Str.1, PO Box 1317, 85741 Garching, Germany\\
  \email{ensslin@mpa-garching.mpg.de~(TAE); volker@mpa-garching.mpg.de~(VS); 
         martin.jubelgas@gmail.com~(MJ)}
  \and
  Canadian Institute for Theoretical Astrophysics, University of Toronto,
  60 St. George Street, Toronto, Ontario, M5S 3H8, Canada\\
  \email{pfrommer@cita.utoronto.ca}
}

\begin{document}
\label{firstpage}


\abstract{
  Cosmic rays (CRs) play a decisive role within our own Galaxy. They
  provide partial pressure support against gravity, they trace past energetic
  events such as supernovae, and they reveal the underlying structure of the
  baryonic matter distribution through their interactions.
  To study the impact
  of CRs on galaxy and cosmic structure formation and evolution, we develop an
  approximative framework for treating dynamical and radiative effects of CRs
  in cosmological simulations. Our guiding principle is to try to find a
  balance between capturing as many physical properties of CR populations as
  possible while at the same time requiring as little extra computational
  resources as possible.
  We approximate the CR spectrum of each fluid element
  by a single power-law, with spatially and temporally varying normalisation,
  low-energy cut-off, and spectral index.  Principles of conservation of
  particle number, energy, and pressure are then used to derive evolution
  equations for the basic variables describing the CR spectrum, both due to
  adiabatic and non-adiabatic processes. The processes considered include
  compression and rarefaction, CR injection via shocks in supernova remnants,
  injection in structure formation shock waves, in-situ re-acceleration of CRs,
  CR spatial diffusion, CR energy losses due to Coulomb interactions,
  ionisation losses, Bremsstrahlung losses, and, finally, hadronic interactions
  with the background gas, including the associated $\gamma$-ray and radio
  emission due to subsequent pion decay.
  We show that the formalism reproduces CR energy densities, pressure, and
  cooling rates with an accuracy of $\sim 10\%$ in steady state conditions
  where CR injection balances cooling. It is therefore a promising formulation
  to allow simulations where CR physics is included.  Finally, we briefly
  discuss how the formalism can be included in Lagrangian simulation methods
  such as the smoothed particle hydrodynamics technique.
Our framework is therefore well suited to be included into numerical simulation
schemes of galaxy and structure formation.
}

\maketitle

\section{Introduction\label{sec:intro}}

\subsection{Motivation\label{sec:motiv}}
The interstellar medium (ISM) of galaxies is highly complex, with an energy
budget composed both of thermal and non-thermal components. Each of the
non-thermal components which are magnetic fields and cosmic rays (CRs) are known to
contribute roughly as much energy and pressure to the ISM as the thermal gas
does, at least in our own Galaxy.  

Magnetic fields permeate the ISM and seem to have ordered and turbulent field
components. Very likely they play an important role in moderating molecular
cloud collapse and star formation. They are the ISM ingredient which couples
otherwise dynamically independent ingredients like the ISM plasma, the CR gas,
and (charged) dust particles into a single, however complex fluid.

\begin{figure}
\resizebox{\hsize}{!}{\includegraphics{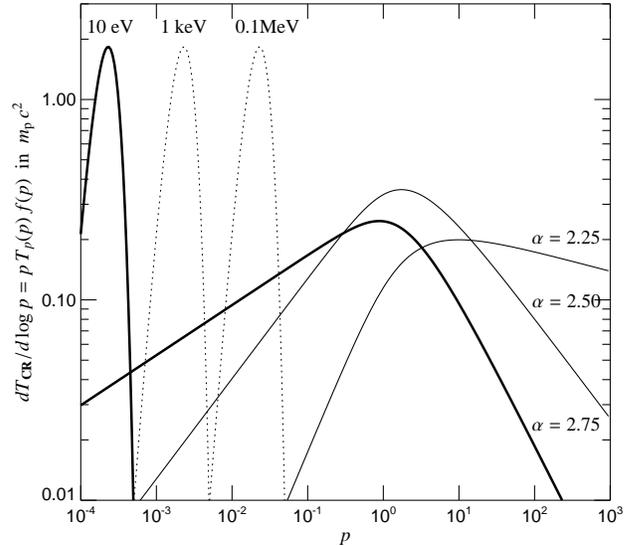}}
\caption{Thermal and power-law (CR) proton spectra for a variety of
  temperatures and spectral indices $\alpha$. No low momentum cutoff was
  imposed onto the CR populations. Displayed is the energy per logarithmic
  momentum interval over dimensionless proton momenta $p = P_{\rm p}/(m_{\rm p}
  \, c^2)$. This form of presentation permits to read-off the dynamically
  important momentum ranges of the populations. All spectra are normalised to
  exhibit the same pressure. The thick lines represents a possible location in
  the hot medium of our Galaxy ($kT = 10$ eV, $\alpha = 2.75$) where thermal
  equals CR pressure. The break in the CR energy distributions is solely due to
  the change in the kinetic energy formulae in the trans-relativistic regime.}
\label{fig:fig_pureSpec}
\end{figure}

CRs behave quite differently compared to the thermal gas. Their equation of
state is softer, they are able to travel actively over macroscopic distances,
and their energy loss time-scales are typically larger than the thermal
ones. Furthermore, roughly half of their energy losses are due to Coulomb and
ionisation interactions and thereby heat the thermal gas. Therefore, CR
populations are an important galactic storage for energy from supernova
explosions, and thereby help to maintain dynamical feedback of star formation
for periods longer than thermal gas physics alone would permit.  The spectral
distribution of CRs is much broader than that of thermal populations, which has
to be taken into account in estimating their dynamical properties. The
dynamical important part of CR spectral distributions in momentum spaces easily
encompasses a few orders of magnitude, whereas thermal distributions appear
nearly mono-energetic on logarithmic scales (see Fig.~\ref{fig:fig_pureSpec}).

Numerical simulations and semi-analytical models of galaxy and large-scale
structure formation neglected the effects of CRs and magnetic fields so far,
despite their dynamical importance. Although there have been attempts to equip
SPH galaxy formation codes with magnetic field descriptions on the MHD level
\citep{1999A&A...348..351D}, a fully dynamical treatment of the CR component
has not yet been attempted due to the very complex CR physics involved. There
have been pioneering efforts to furnish grid-based MHD codes with a diffusive
CR component in order to study isolated effects like Parker instabilities
\citep{2004ApJ...607..828K, 2003A&A...412..331H} or the injection of CRs into
the wider IGM \citep{2001ApJ...559...59M, 2003ApJ...593..599R,
2003JKAS...36..105R, 2004JKAS...37..477R} and ISM
\citep{2006MNRAS.373..643S}. However, these codes are not suited for galaxy
formation simulations in a cosmological setting due to the missing adaptive
resolution capability, and the lack of a treatment of cosmological components
such as dark matter and stellar populations. There have also been numerical
implementations of discretised CR energy spectra on top of grid-based
cosmological simulations~\citep{2001CoPhC.141...17M}, but these implementations
neglected the hydrodynamic pressure of the CR component.  In addition, the
amount of computer resources required for these models in terms of memory and
CPU-time is substantial.  This renders this approach unattractive for
self-consistent simulations of galaxy formation and inhibits the inclusion of
CRs into cosmological simulations of structure formation, where it is not clear
ab initio if CRs are crucial or not.

An accurate description of CRs should follow the evolution of the spectral
energy distribution of CRs as a function of time and space, and keep track of
their dynamical, non-linear coupling with the hydrodynamics.  In order to
allow the inclusion of CRs and their effects into numerical simulations and
semi-analytic descriptions of galaxy formation, we develop a simplified
description of the CR dynamics and physics. We seek a compromise between two
opposite requirements, namely: (i) to capture as many physical properties and
peculiarities of CR populations as possible, and (ii) to require as little
extra computational resources as possible.  In our model, the emphasis is
given to the dynamical impact of CRs on hydrodynamics, and not on an accurate
spectral representation of the CRs. The guiding principles are energy,
pressure, and particle number conservations, as well as adiabatic invariants.
Non-adiabatic processes will be mapped onto modifications of these principles.

The goal of the formalism is to permit explorative studies investigating under
which circumstances CRs are likely to affect galactic or cosmic structure
formation processes. The approach is phenomenological, which implies that any
adopted model parameter like CR injection efficiency or diffusivity has to be
verified by independent means before definite conclusions may be drawn, if the
conclusions depend strongly on those assumptions. Nevertheless, it is plausible
that in many cases the overall physical picture will not sensitively depend on
the details of the CR description, but merely on the presence of CRs. An
implementation of the formalism in a cosmic structure formation code will
therefore allow an identification of regimes in which CRs are likely to have a
significant impact on the properties of galaxies and galaxy clusters. We want
to stress that including ``only'' an approximate description of CR dynamics
into galaxy and structure formation calculations should be regarded as a
significant improvement compared to the current situation where the CRs are
ignored completely.

In this paper, we give the theoretical basis and justification of our
formalism.  An implementation and first applications of our formalism are
described by two companion papers. \citet{Jubelgas} describe the implementation
into the GADGET code and present first results on galaxy formation including
CRs. \citet{2006MNRAS.367..113P} develop a SPH Mach number measuring scheme, which is a
necessary prerequisite to study CRs injected at accretion shock waves in the
cosmic large-scale structure.

\subsection{Approximations and assumptions\label{sec:assump}}

A number of simplifying assumptions are necessary in order to reduce the large
number of degrees of freedom of a CR population before it can be included in a
numerical tractable scheme. For transparency, we list the main assumptions here:
\begin{enumerate}
\item {\bf Protons dominate the CR population:} We only model the dominant CR
  proton population, assuming that the presence of $\alpha$-particles and
  heavier ions would not change the dynamical picture. In contrast to heavier
  ions, $\alpha$-particles carry a significant fraction of the total CR
  energy. Nevertheless, our assumption is a reasonable approximation, since the
  energy density of $\alpha$-particles can be absorbed into the proton
  spectrum. As a coarse approximation, a GeV energy $\alpha$-particle can be
  regarded as an ensemble of four individual nucleons traveling together due
  to the relatively weak MeV nuclear binding energies compared to the kinetic
  energy of relativistic protons. For hadronic interactions, the fact that the
  four nucleons are bound is of minor importance. Since Coulomb cooling and
  ionisation losses are proportional to the square of the nucleus' charge, each
  of the four nucleons of the $\alpha$-particle is experiencing a loss of
  kinetic energy which is identical to the loss that a free CR proton with
  exactly the same specific energy would experience. Furthermore, due to their
  low galactic energy density, but also due to their more complex radiative
  losses, we ignore CR electron populations. Only a description of
  quasi-equilibrium secondary electrons resulting from hadronic CR-gas
  interactions is provided, which has some restricted applicability in galaxy
  cluster physics.
\item {\bf A momentum power-law is a typical spectrum:} We assume that the CR
  momentum spectrum can be well described by a power-law with spectral index
  $\alpha$.  This is not only consistent with the observation of Galactic CRs
  ($\alpha \approx 2.75$) but is also predicted by many CR acceleration and
  diffusion models, which frequently give momentum power-law spectra
  \citep[see][ for a review]{2002cra..book.....S}. Especially CR injection at
  shock waves in the test-particle limit predicts power-law spectra in momentum
  space.  The dynamically relevant physical quantities of the CR population are
  its kinetic energy density $\eps_\CR$, the average energy $T_\CR =
  \eps_\CR/n_\CR$, and the pressure $P_\CR$. For an assumed power-law spectrum
  with steep spectral index $\alpha$ at high CR momenta, these two quantities
  are completely determined by a specification of the normalisation constant
  $C$ and the low-momentum cutoff $q$ of the power-law. No high-momentum cutoff
  of the spectrum is considered, since for a sufficiently steep spectrum
  ($\alpha>2$), the high-energy range is dynamically unimportant. Instead, the
  dynamics is dominated by particles with momenta closest to $q\,\mp\,c$ (the
  particles at the lower cutoff), or around $\mp\,c$, whichever is larger
  (and the latter only if $\alpha <3$, otherwise the particles at
  $q\,\mp\,c$ dominate always).
\item {\bf Energy, particle number, and pressure are the relevant CR
  variables:} The many degrees of freedom a CR spectrum have to be
  reduced to simplified evolution equations for the spectral normalisation
  constant $C$, for the cutoff $q$, and eventually for the spectral index
  $\alpha$.  The mapping of the full spectral dynamics on this restricted set
  of variables is not unique. However, physical intuition tells us that the
  most relevant quantities, which should be reproduced in the formalism with
  highest accuracy, are the CR energy density, the number density, and the
  pressure. CR pressure, which is very similar to the CR energy density, will
  only be used if a varying spectral index is chosen. Therefore, whenever a
  physical effect modifies a CR spectrum, the changes on $C$, $q$, and
  eventually $\alpha$ will be calculated by taking energy, particle and
  eventually pressure conservation into account. Whenever we have to make a
  choice we will favour accuracy in CR energy over accuracy in CR number
  density. This choice is motivated by our aim to study the dynamical effects of a CR
  population on their host environment. Spectral details of the CR populations,
  which are not the focus of our formalism, are therefore treated very
  coarsely.
\end{enumerate}
We present two models for the description of the CR population that differ in
the degree of their complexity. In the simpler model, we assume the CR spectral
index $\alpha$ to be the same everywhere, while in the more complex model we
allow it to vary in space and time. The model with the constant spectral index
suppresses a number of CR phenomena like spectral steepening due to transport
effects, or spectral flattening due to fresh particle injection.  However,
these effects are probably not essential for a first order description of the
global dynamical impact of CRs.

\subsection{Captured physics\label{sec:CapturedPhys}}

Our framework is set up to accommodate a number of essential physical processes
of CR gases, like particle acceleration, diffusion, and particle interactions.
The different components of our framework are of different relevance in the
various astrophysical environments:
\begin{enumerate}
  \item CR energy is an important energy storage within galactic
  metabolisms. In our Galaxy, the CR energy density is comparable to the thermal
  one. 
  \item CR pressure is suspected to play an important role in forming galactic
  winds.
  \item Adiabatic energy losses and gains of CRs are the mechanical way to exchange
  energy with the thermal gas, and therefore relevant for CR driven winds or
  radio bubbles in cluster atmospheres. 
  \item Shock acceleration is believed to be an important source of
  CRs. In the galactic context, supernova shock waves should dominate the
  energy injection, in galaxy clusters, the cluster merger and accretion shock
  waves may be more efficient.
  \item The role of CR diffusion is actually not always clear. Diffusive escape times
  in galaxy clusters exceed the Hubble time for GeV particles, whereas we know
  that CRs escape from our own galaxy on time-scales of $10^7$ years, which
  certainly involves diffusive CR transport.
  \item In situ CR re-acceleration by plasma waves may energise relativistic
  electrons in several astrophysical environments.  This process may be
  important for cluster radio halos but also in several other astrophysical
  situations (Galaxy, Sun,..).\label{point:RAcc}
  \item Coulomb and ionisation energy losses of the CRs are heating the gas,
  thereby possibly maintaining a minimum temperature and ionisation in dense
  molecular clouds.
  \item Bremsstrahlung from CR protons is not an important effect -- to our
  knowledge --, neither for the CR cooling, nor for the production of
  high-energy photons.
  \item Hadronic interactions is the only efficient non-adiabatic energy loss
  process of ultra-relativistic protons. Their secondaries (photons, electrons)
  are important tracers of CR populations. Gamma rays from such interactions
  are observed from our Galaxy. In galaxy clusters, synchrotron emission of
  secondary electrons is proposed as a possible origin of the giant radio
  halos (among other possibilities, see point \ref{point:RAcc}).
\end{enumerate}

Thus, these effects should allow realistic studies of the impact of a
variety of physical processes of CRs on galaxies, clusters of galaxies, and on
the large-scale intergalactic medium, including:
\begin{enumerate}
\item hydrodynamical effects of CRs,
\item CR injection by diffusive shock acceleration,
\item in-situ re-acceleration by plasma waves,
\item non-local feedback from CR injection due to CR diffusion,
\item CR modified shock structures, 
\item heating of cold gas by CRs,
\item CR driven galactic winds,
\item Parker instabilities of spiral galaxy disks,
\item rough morphology of gamma ray emission,
\item rough morphology of radio emission due to secondary electrons.
\end{enumerate}
We like to point out that our approach is a phenomenological one which does not
provide first principle descriptions of the above physical effects. It is
also not always accurate, especially when spectral effects become relevant and
deviations from power-law spectra are expected, e.g. in the case of non-linear
shock acceleration.  Although our formalism captures many properties only
approximately, it allows to study many astrophysical consequences in greater
detail than before and also permits to study the influence of
the adopted CR parameters like injection efficiency, diffusivity, etc. on
observable quantities.

\subsection{Structure of paper}\label{sec:struct}
Our basic formalism is outlined in Sect.~\ref{sec:formalism}, in which the
approximative description of the CR gas is introduced and its adiabatic
evolution is described (Sect.~\ref{sec:bascis}). A model with a constant
spectral index for the power-law description of the CR population is given in
Sect.~\ref{sec:constalpha} while the more general approach with a spatially and
temporally varying spectral index is presented in
Sect.~\ref{sec:variablealpha}.  Sect.~\ref{sec:nadiab} contains the technical
description of the various non-adiabatic processes: CR injection via shocks
waves of structure formation (Sect.~\ref{sec:CRinjStFo}), and of supernova
remnants (Sect.~\ref{sec:CRinj}); CR spatial diffusion
(Sect.~\ref{sec:diffusion}); in-situ re-acceleration of CRs
(Sect.~\ref{sec:insitu}); CR energy losses due to Coulomb interactions
(Sect.~\ref{sec:Coulomb}), ionisation losses (Sect.~\ref{sec:Ionisation}),
Bremsstrahlung (Sect.~\ref{sec:Brems}), and hadronic interactions with the
background gas (Sect.~\ref{sec:caloss}), including the associated
$\gamma$-radiation from the $\pi^0$-decays (Sect.~\ref{sec:gammas}) and the
radio emission of the electrons and positrons resulting from $\pi^\pm$-decays
(Sect.~\ref{sec:hadron}). Steady state CR spectra, for which injection and
cooling balance each other, and freely cooling CR spectra are calculated
exactly and with our formalism are compared in
Sect.~\ref{sec:tst}. Section~\ref{sec:sph} describes how the formalism can be
included into a simulation code based on smoothed particle hydrodynamics
(SPH). Finally, we give our conclusions in Sect.~\ref{seq:concl}.

\section{Formalism}\label{sec:formalism}

In this section, we develop a description of a CR population in a volume
element which is comoving with the background fluid. This Lagrangian
perspective results in a significant simplification, since the advective
transport processes are fully characterised by a description of the effects of
adiabatic volume changes. We will introduce convenient, adiabatically
invariant variables which are constant in time in the absence of non-adiabatic
processes. When non-adiabatic processes are included, they can be expressed in
terms of evolutionary equations for the adiabatically invariant variables.
The chosen formalism is well suited for the implementation in Lagrangian
numerical simulation codes, as will be discussed in more detail in
Sect.~\ref{sec:sph} for the example of SPH codes.

\subsection{Basic cosmic ray variables}
\label{sec:bascis}

Since we only consider CR protons, which are at least in our Galaxy the
dominant CR species, it is convenient to introduce the dimensionless momentum
$p = P_\p/(\mp\,c)$.  We assume that the differential particle momentum
spectrum per volume element can be approximated by a single power-law above
the minimum momentum $q$:
\begin{equation}
\label{eq:spec1}
f(p) = \frac{\dd N}{\dd p\,\dd V} = C \, p^{-\alpha}\,
\theta(p- q) ,
\end{equation}
where $\theta(x)$ denotes the Heaviside step function. Note that we use an
effective one-dimensional distribution function $f(p)\equiv 4\upi p^2
f^{(3)}(p)$.  The differential CR spectrum can vary spatially and temporally
(although for brevity we suppress this in our notation) through the spatial
dependence of the normalisation $C=C(\vecbf{x},t)$ and the cutoff $q =
q(\vecbf{x},t)$. In the simpler of our two models, we assume the CR spectral
index $\alpha$ to be constant in space and time, while the more complex model
allows for a spatial and temporal variation of $\alpha$ as well.

Adiabatic compression or expansion leaves the phase-space density of the CR
population invariant, leading to a momentum shift according to $p_0 \rightarrow
p = (\rho/\rho_0)^{1/3}\, p_0$ for a change in density from $\rho_0$ to $\rho$.
Since this is fully reversible, it is useful to introduce an invariant cutoff
$q_0$ and a normalisation $C_0$ which describe the CR population via
Eqn.~(\ref{eq:spec1}) if the gas is adiabatically compressed or expanded
relative to the reference density $\rho_0$. The actual parameters of the
spectrum are then given by
\begin{equation}
\label{eq:adiabatic}
q(\rho) = \left({\rho}/{\rho_0} \right)^{\frac{1}{3}}\, q_0\;\;
\mbox{and}\;\; C(\rho) = \left({\rho}/{\rho_0}
\right)^{\frac{\alpha+2}{3}}\, C_{0}.
\end{equation}
The variables $q_0$ and $C_0$ are hence a convenient choice for a Lagrangian
description of the ISM or the intra-cluster medium (ICM).

The CR number density is given by
\begin{equation}
\label{eq:ncr}
n_{\CR} = \int_0^\infty \!\!\!\! \!\! \dd p\, f(p) =
\frac{C\, q^{1-\alpha}}{\alpha-1} =
\frac{C_0\,q_0^{1-\alpha} }{\alpha-1}\, \frac{\rho}{\rho_0}\,,
\end{equation}
provided $\alpha >1$.
The kinetic energy density of the CR population is
\begin{eqnarray}
\label{eq:epscr}
\eps_\CR &=& \int_0^\infty \!\!\!\! \!\! \dd p\, f(p) \,T_{\rm
p}(p)=\frac{C\, \mp\,c^2}{\alpha-1} \, \times
\nonumber \\
&& \left[\frac{1}{2}
\, \B_{\frac{1}{1+q^2}} \left(
\frac{\alpha-2}{2},\frac{3-\alpha}{2}\right) + q^{1-\alpha}
\left(\sqrt{1+q^2}-1 \right) \right] \,,
\end{eqnarray}
where $T_\p(p) = (\sqrt{1+p^2} -1)\, \mp\,c^2$ is the kinetic energy of a
proton with momentum $p$. $\B_x(a,b)$ denotes the incomplete Beta-function,
and $\alpha>2$ is assumed. The average CR kinetic energy $T_\CR =
\eps_\CR/n_\CR$ is therefore
\begin{equation}
\label{eq:Tcr}
T_\CR = \left[\frac{q^{\alpha-1}}{2} \, \B_{\frac{1}{1+q^2}}
\left( \frac{\alpha-2}{2},\frac{3-\alpha}{2}\right) + \sqrt{1+q^2}-1
\right]\, {\mp\,c^2}.
\end{equation}
The CR pressure is
\begin{equation}
\label{eq:Pcr}
P_\CR = \frac{\mp c^2}{3}\,\int_0^\infty \!\!\!\! \!\! \dd p\, f(p)
\,\beta\,p  =\frac{C\,\mp c^2}{6} \, 
\B_{\frac{1}{1+q^2}} \left( \frac{\alpha-2}{2},\frac{3-\alpha}{2}
\right) ,
\end{equation}
where $\beta = \vel/c = p/\sqrt{1+p^2}$ is the dimensionless velocity of the
CR particle.  The CR population can hydrodynamically be described by an
isotropic pressure component as long as the CRs are coupled to the thermal gas
by small scale chaotic magnetic fields. Note, that for $2<\alpha<3$ the kinetic
energy density and pressure of the CR populations are well defined for the
limit $q\rightarrow 0$, although the total CR number density diverges.

\begin{figure}
\resizebox{\hsize}{!}{\includegraphics{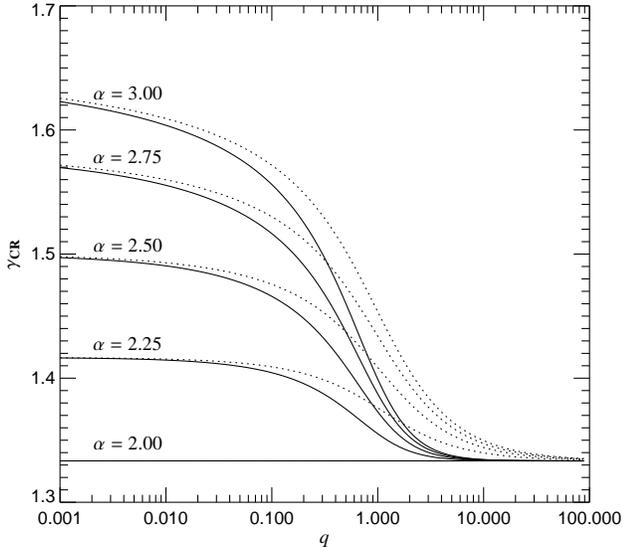}}
\caption{Adiabatic index $\gamma_{\CR}$ of a CR population as a function of the
  lower momentum cutoff $q$ for a variety of spectral indices $\alpha$ (solid
  lines). Furthermore, the pseudo-adiabatic index $\gamma_{\CR}^* = 1 +
  P_\CR/\eps_\CR$ relating the total CR pressure and energy density is also
  displayed (dotted lines). Only for a polytropic gas behaviour we have
  $\gamma_{\CR} = \gamma_{\CR}^*$. This applies approximately for CR
  populations which are either completely non- or completely
  ultra-relativistic. However, in the here important case of trans-relativistic
  CR populations polytropic equations of state can not be assumed.}   
\label{fig:fig_gamma}
\end{figure}

The local adiabatic exponent of the CR population is defined by 
\begin{equation}
\label{eq:gammaCR}
\gamma_\CR \equiv \left.\frac{\dd \log P_\CR}{\dd \log \rho}\right|_S,
\end{equation}
where the derivative has to be taken at constant entropy $S$. Using
Eqns.~(\ref{eq:adiabatic}) and (\ref{eq:Pcr}), we obtain for the CR adiabatic
exponent
\begin{eqnarray}
\label{eq:gammaCR2}
\gamma_\CR &=& \frac{\rho}{P_\CR}
\left(\frac{\partial P_\CR}{\partial C}\frac{\partial C}{\partial \rho}
+ \frac{\partial P_\CR}{\partial q}\frac{\partial q}{\partial \rho}\right) \nonumber\\
&=& \frac{\alpha + 2}{3} - \frac{2}{3}\,
q^{2-\alpha}\, \beta(q)\,
\left[\B_{\frac{1}{1+q^2}} 
 \left( \frac{\alpha-2}{2},\frac{3-\alpha}{2}\right) \right]^{-1}.
\end{eqnarray}
Note that  the CR adiabatic
exponent is time dependent due to its dependence on the lower cutoff of the CR
population, $q$, which is shown in Fig.~\ref{fig:fig_gamma}.  For a mixture of
thermal and CR gas, it is appropriate to define an effective adiabatic index by 
\begin{equation}
\label{eq:gammaeff}
\gamma_\rmn{eff} \equiv \left.\frac{\dd \log (P_\rmn{th} + P_\CR)}
{\dd \log \rho}\right|_S = 
\frac{\gamma_\rmn{th}\, P_\rmn{th} + \gamma_\CR\, P_\CR}
{P_\rmn{th} + P_\CR}.
\end{equation}

\subsection{Formalism with a constant spectral index}
\label{sec:constalpha}

The spectrum of the CR population within a given fluid element is shaped by a
number of different physical processes, such as particle injection and escape,
continuous and hadronic energy losses, or re-acceleration. While all of
these processes leave a characteristic signature in the CR spectrum, in the
framework of our simplified model we have to describe their effects in terms
of the two dynamical variables, $C$ and $q$ (or $C_0$ and $q_0$). In order to
approximate the key features of the real CR dynamics with our 
description, we have to make a proper choice for how to modify our variables
by the different processes. The guiding principle for this are energy and
particle number conservation.

Consider a non-adiabatic process leading to a change $\dd n_\CR$ in the number
density and a change $\dd\eps_\CR$ in the energy density of the particles
during an infinitesimal time interval $\dd t$. The implied change in
$(C,q)$ is then given by
\begin{equation}
\left(
\begin{array}{c}
\dd C\\ \dd q
\end{array}
\right)
= 
\left(
\begin{array}{cc}
\partial C/\partial n_\CR & \partial C/\partial \eps_\CR\\
\partial q/\partial n_\CR & \partial q/\partial \eps_\CR
\end{array}
\right)
\left(
\begin{array}{c}
\dd n_\CR\\ \dd\eps_\CR
\end{array}
\right).
\end{equation}
The inverse of the Jacobian can be straight forwardly calculated using the
definitions in Eqns. (\ref{eq:ncr}) and (\ref{eq:epscr}) which can easily be
inverted to yield
\begin{eqnarray}
\dd C &=& C \,\frac{\dd\eps_\CR - T_\p(q)\,\dd n_\CR}{\eps_\CR
- T_\p(q)\, n_\CR },\\
\dd q &=& \frac{q}{\alpha -1} \,\frac{\dd\eps_\CR - T_\CR\,\dd n_\CR}{\eps_\CR
- T_\p(q)\, n_\CR }\,.
\end{eqnarray}
These relations are reasonable, which can also be verified by the following
gedanken experiments: if a process increases $\eps_\CR$ and $n_\CR$
simultaneously by the same factor $1+\delta$, so that $\dd\eps_\CR/\eps_\CR =
\dd n_\CR/n_\CR = \delta$, one gets only a change in the normalisation ($\dd C
= \delta \,C$), but not in the cutoff ($\dd q =0$), as it should be. If one
adds an infinitesimal amount of particles $\dd n_\CR$ with exactly the kinetic
energy of the cutoff $T_\p(q)$, so that $\dd\eps_\CR = T_\p(q) \,\dd n_\CR$,
the normalisation is unchanged ($\dd C =0$), but the cutoff is lowered ($\dd q
= -q\,\dd n_\CR/[(\alpha-1)\,n_\CR] \Rightarrow n_\CR \propto q^{1-\alpha}$),
again as expected.

The adiabatically invariant variables change according to
\begin{eqnarray}
\label{eq:dC0}
\dd C_0 &=& \left(\frac{\rho}{\rho_0} \right)^{-\frac{\alpha+2}{3}}\, \dd C = C_0
\,\frac{\dd\eps_\CR - T_\p(q)\,\dd n_\CR}{\eps_\CR - T_\p(q)\,
n_\CR }\,,\\
\label{eq:dq0}
\dd q_0 &=& \left(\frac{\rho}{\rho_0} \right)^{-\frac{1}{3}}\, \dd q
= \frac{q_0}{\alpha -1} \,\frac{\dd\eps_\CR - T_\CR\,\dd n_\CR}{\eps_\CR
- T_\p(q)\, n_\CR }\,,
\end{eqnarray}
where we used a notation which is mixed in the variant and invariant
variables, for convenience.  Ways to numerically implement the evolution of
$C_0$ and $q_0$ are discussed in Appendix~\ref{sec:updateconst}.

\subsection{Formalism with a variable spectral index}
\label{sec:variablealpha}

The modeling of certain effects of CR physics such as spectral steepening due
to transport effects, or spectral flattening due to fresh particle injection
requires a more elaborate description of the CR population.  Thus, we
generalise the above formalism by allowing the spectral index to vary in space
and time as well. To this end, we now consider the CR population to be
described by three dynamical variables $C$, $q$, and $\alpha$. In addition to
energy and particle number conservation, we invoke pressure conservation.  As a
word of caution, we note that the resulting formalism is considerably more
demanding in terms of computational resources than the formulation with a
constant spectral index. We hence suggest that this description is only used
if the problem under consideration requires this additional information.

For notational purposes, we define a three-vector with the dynamical CR
variables as $\vecbf{a} \equiv (C,q,\alpha)^\rmn{T}$, where the exponent
denotes the transpose of the vector.  Similarly, we define a vector of
hydrodynamic CR quantities $\vecbf{h} \equiv (n_\CR,\eps_\CR,P_\CR)^\rmn{T}$.
Consider a non-adiabatic process leading to an infinitesimal change $\dd
n_\CR$ in the number density, $\dd\eps_\CR$ in the energy density, and $\dd
P_\CR$ in the pressure of the particles during an infinitesimal time interval
$\dd t$. This implies a change in $\vecbf{a}=(C,q,\alpha)^\rmn{T}$ 
given by
\begin{equation}
  \label{eq:variablealpha}
  \dd a_i = \sum_{j=1}^3 \frac{\partial a_i}{\partial h_j} \dd h_j
  \quad\Leftrightarrow\quad
  \dd \vecbf{a} = \bf{\sf A}\, \dd \vecbf{h}.
\end{equation}
Since the entries of the Jacobian $\bf{\sf A}$ cannot be straightforwardly
obtained, we propose to compute them by inverting the inverse of the Jacobian
$\bf{\sf A}^{-1}$. Once this is achieved, the changes $\dd \vecbf{a}$ in the dynamical CR
variables  can be easily obtained. The inverse of the Jacobian
is given by
\begin{eqnarray}
  \label{eq:invJacobian}
  \bf{\sf A}^{-1} &=&\left( 
    \begin{array}{lll}
      \frac{\dps n_\CR}{\dps C} & -\frac{\dps \alpha-1}{\dps q} n_\CR 
      & \frac{\dps \partial n_\CR}{\dps \partial \alpha} \\
      \rule{0mm}{5mm} 
      \frac{\dps \eps_\CR}{\dps C} 
      & -\frac{\dps \alpha-1}{\dps q}\, n_\CR\, T_\p(q) 
      & \frac{\dps \partial \eps_\CR}{\dps \partial \alpha} \\
      \rule{0mm}{7mm} 
      \frac{\dps P_\CR}{\dps C} 
      & -\frac{\dps m_\p\, c^2\, (\alpha-1)}{\dps 3}\,\beta(q)\, n_\CR 
      & \frac{\dps \partial P_\CR}{\dps \partial \alpha} \\
    \end{array} 
  \right) ,
\end{eqnarray}
where the last column can be  explicitly expressed as
\begin{eqnarray}
  \frac{\partial n_\CR}{\partial \alpha} &=&
  -\frac{n_\CR}{\alpha-1}[1+(\alpha-1)\,\ln q], \\
  \frac{\partial \eps_\CR}{\partial \alpha} &=&
  -\int_0^\infty \!\!\!\!\! \dd p\, T_\p(p)\, f(p)\,\ln p , \\
  \frac{\partial P_\CR}{\partial \alpha} &=&
  -\frac{m_\p\, c^2}{3}\int_0^\infty \!\!\!\!\!  \dd p\, p\, \beta(p)\, f(p)\,\ln p .
\end{eqnarray}
We discuss ways to numerically implement the time evolution of $C_0$, $q_0$,
and $\alpha$ in Appendix~\ref{sec:updatevariable}. 

A note of caution should be in order. For ultra-relativistic CR populations
the formalism with varying spectral index becomes degenerate, since in this
regimes the equation of state is independent of the spectral index (see
Fig. \ref{fig:fig_gamma}). However, as can be seen in Sect. \ref{sec:tst}, the
dynamics of CR cooling always pushes the low-energy cutoff into the
trans-relativistic regime, where this formalism should work.

\section{Non-adiabatic processes}\label{sec:nadiab}

In the following, we discuss various non-adiabatic source and sink processes
of the CR population. In each subsection, we outline the physical motivation
of each process and describe its effects in terms of a change in energy and
number density. Afterwards, we generalise to the more complex case of a
spatially and temporally varying CR spectral index.

\subsection{CR shock acceleration}\label{sec:acceleration}
In this section, we discuss CR acceleration processes at shock waves due to
gas accretion and galaxy mergers, using the framework of {\em diffusive shock
  acceleration}.  Our description is a modification of the approach of
\citet{2001CoPhC.141...17M}.  The shock surface separates two regions: the
{\em upstream regime} defines the region in front of the shock which is
causally unconnected for super sonic shock waves, whereas the {\em downstream
  regime} lies in the wake of the shock wave.  The shock front itself is the
region in which the mean plasma velocity changes rapidly on small scales,
governed by plasma physics. In the rest frame of the shock, particles are
impinging onto the shock surface at a rate per unit area of $\rho_2 \vel_2 =
\rho_1 \vel_1$. Here $\vel_1$ and $\vel_2$ give the plasma velocities
(relative to the shock's rest frame) in the upstream and downstream regimes of
the shock, respectively. The corresponding mass densities are denoted by
$\rho_1$ and $\rho_2$.

We assume that after passing though the shock front most of the gas thermalises
to a Maxwell-Boltzmann distribution with characteristic post-shock temperature
$T_2$:
\begin{equation}
  \label{eq:MBdistibution}
  f_\rmn{th2}(p) = 4\upi\, n_\rmn{th}\,
  \left(\frac{m_\p c^2}{2 \upi\, k T_2}\right)^{3/2}\! p^2
  \exp\left(-\frac{m_\p c^2\,p^2 }{2 \,k T_2}\right),
\end{equation}
where the number density of particles of the thermal distribution in the
downstream regime, $n_\rmn{th}=n_2$, as well as $T_2$ can be inferred by means
of the mass, momentum, and energy conservation laws at the shock surface for a
gas composed of CRs and thermal constituents.  In
a companion paper, we describe a formalism for instantaneously and
self-consistently inferring the shock strength and all other relevant
quantities in the downstream regime of the shock within the framework of SPH
simulations \citep{2006MNRAS.367..113P}.  Assuming that a fraction of the
thermalised particles experience stochastic shock acceleration by diffusing
back and forth over the shock front, the test particle theory of diffusive
shock acceleration predicts a resulting CR power-law distribution in momentum
space.  Within our model, this CR injection mechanism can be treated as an
instantaneous process.

For a particle in the downstream region of the shock to return upstream it is
necessary to meet two requirements. The particle's effective velocity
component parallel to the shock normal has to be larger than the velocity of
the shock wave, and secondly, its energy has to be large enough to escape the
``trapping'' process by Alfv\'en that are generated in the downstream
turbulence \citep{1995A&A...300..605M,1998AdSpR..21..551M}.  Thus, only
particles of the high-energy tail of the distribution are able to return to
the upstream shock regime in order to become accelerated. The complicated
detailed physical processes of the specific underlying acceleration mechanism
are conveniently compressed into a few parameters
\citep{1993ApJ...402..560J,1994APh.....2..215B,1995ApJ...447..944K}, one of
which defines the momentum threshold for the particles of the thermal
distribution to be accelerated,
\begin{equation}
  \label{eq:qinj}
  q_\rmn{inj} = x_\rmn{inj} p_\rmn{th} = 
  x_\rmn{inj} \sqrt{\frac{2 \,k T_2}{m_\p c^2}}.
\end{equation}
Since Coulomb and ionisation losses efficiently modify the low energy part of
the injected CR spectrum, we propose to follow the recipe presented at the end
of Sect.~\ref{sec:CRinj}, i.e.~to simply increase the low energy spectral break
of the actually injected spectrum without changing the normalisation of the CR
spectrum.

In the linear regime of CR acceleration, the thermal distribution joins in a
smooth manner into the resulting CR power-law distribution at $q_\rmn{inj}$ so
that $x_\rmn{inj}$ represents the only parameter in our simplified diffusive shock
acceleration model,
\begin{equation}
  \label{eq:finj}
  f_\rmn{CR,lin}(p) = 
  f_\rmn{th}(q_\rmn{inj}) \left(\frac{p}{q_\rmn{inj}}\right)^{-\alpha_\rmn{inj}} 
  \theta(p-q_\rmn{inj}).
\end{equation}
Fixing the normalisation of the injected CR spectrum by this continuity
condition automatically determines $C_\rmn{inj}$ which depends on the second
adiabatic invariant.  The slope of the injected CR spectrum is given by
\begin{equation}
  \label{eq:ainj}
  \alpha_\rmn{inj} = \frac{r + 2}{r - 1}, \quad\mbox{where}\quad
  r = \frac{\rho_2}{\rho_1} = \frac{\vel_1}{\vel_2} 
\end{equation}
denotes the shock compression ratio \citep{1978MNRAS.182..147B,
  1978MNRAS.182..443B, 1983SSRv...36...57D}.  In the linear regime, the number
density of injected CR particles is given by
\begin{equation}
  \label{eq:ninj}
  \Delta n_\rmn{CR,lin} = \int_0^\infty \!\!\!\!\! \dd p\, f_\rmn{CR,lin}(p)
  = f_\rmn{th}(q_\rmn{inj})\, \frac{q_\rmn{inj}}{\alpha_\rmn{inj}-1}.
\end{equation}
This enables us to infer the particle injection efficiency which is a measure
of the fraction of downstream thermal gas particles which experience diffusive
shock acceleration,
\begin{equation}
  \label{eq:eta}
  \eta_\rmn{CR,lin}\equiv\frac{\Delta n_\rmn{CR,lin}}{n_\rmn{th}} = 
  \frac{4}{\sqrt{\upi}}\,\frac{x_\rmn{inj}^3}{\alpha_\rmn{inj}-1}\,
  \rmn{e}^{-x_\rmn{inj}^2}.
\end{equation}
The particle injection efficiency is independent of the downstream post-shock
temperature $T_2$.  These considerations allow us to infer the dynamically
relevant injected CR energy density in the linear regime:
\begin{equation}
  \Delta\eps_\rmn{CR,lin}  =
  \eta_\rmn{CR,lin}T_\CR(\alpha_\rmn{inj},q_\rmn{inj})\,n_\rmn{th}(T_2). 
\end{equation}
In our description, the CR energy injection efficiency in the linear regime is
defined to be the energy density ratio of freshly injected CRs to the total
dissipated energy density in the downstream regime,
\begin{equation}
  \zeta_\rmn{lin} =
  \frac{\Delta\eps_\rmn{CR,lin}}{\Delta\eps_\rmn{diss}},
   \quad\mbox{where}\quad
  \Delta\eps_\rmn{diss} = \eps_\rmn{th2} - \eps_\rmn{th1} r^\gamma.
\end{equation}
The dissipated energy density in the downstream regime,
$\Delta\eps_\rmn{diss}$, is given by the difference of the thermal energy
densities in the pre- and post-shock regimes, corrected for the
contribution of the adiabatic part of the energy increase due to the
compression of the gas over the shock.

In order to obey energy conservation as well as the equations of the linear
theory of diffusive shock acceleration, $\zeta_\rmn{lin}$ has to fulfill a
boundary condition which ensures that the dynamical pressure exerted by CRs is
smaller than the ram pressure $\rho_1 \vel_1^2$ of the flow, yielding
\begin{equation}
  \label{eq:CRtoRam}
  \frac{P_\CR}{\rho_1 \vel_1^2} = 
  \frac{(\alpha-1)\, c^2\, \eta_\rmn{CR,lin}}{6\, \vel_1 \vel_2}
   q_\rmn{inj}^{\alpha-1} 
  \B_{\frac{1}{1+q_\rmn{inj}^2}} 
  \left( \frac{\alpha-2}{2},\frac{3-\alpha}{2}\right)< 1,
\end{equation}
where $\alpha = \alpha_\rmn{inj}$. 

In typical applications like cosmological SPH simulations, this boundary
condition can be substantially simplified. To this end, we propose the
following modification of the CR energy injection efficiency in order to
account for the saturation effect at high values of the Mach number:
\begin{equation}
\label{eq:saturation}
\zeta_\rmn{inj} = \left[1 - \exp\left(-\frac{\zeta_\rmn{lin}}
  {\zeta_\rmn{max}}\right)\right]\,\zeta_\rmn{max}.
\end{equation}
Numerical studies of shock acceleration suggest a value of $\zeta_\rmn{max}
\simeq 0.5$ for the limiting case of the CR energy injection efficiency
\citep{2003ApJ...593..599R}.  One can then infer the injected CR energy
density in terms of the energy injection efficiency of diffusive shock
acceleration processes,
\begin{equation}
  \label{eq:zeta}
\Delta \eps_\rmn{CR,inj} = \zeta_\rmn{inj} \Delta \eps_\rmn{diss}.
\end{equation}
We note that nonlinear effects, in the form of back-reactions of the
accelerated particles upon the shock, change the expectations with respect to
the linear case. These effects are expected to be important well before the
limit given by Eqn.~(\ref{eq:saturation}) \citep[see][ and many
others]{1979ApJ...229..419E, 1981ApJ...248..344D, 1982A&A...111..317A,
2000ApJ...533L.171M, 2002APh....16..429B, 2005MNRAS.361..907B,
2005ApJ...620...44K}. 

\del{We note, that in case of simulations including CR-diffusion, shock
injected CRs may diffuse upstream within the simulation and build up a
CR-precursor of the shock wave. The CR pressure gradient of the precursor will
will adiabatically heat the incoming plasma before it reaches the shock wave
and thereby modify the shock structure, just as it should happen in a real
non-linear modified shock acceleration scenario. Thus, such non-linear
modifications may occur automatically within the simulation if CR diffusion is
permitted. However, to which extent the coarser spatial and spectral
description in such a simulation will reproduce the results of analytical
studies of non-linear particle acceleration has to be investigated separately.}

The average kinetic energy of $T_\CR(\alpha_\rmn{inj}, q_\rmn{inj})$
of an injection power-law spectrum with CR spectral index $\alpha_\rmn{inj}$ is
given by Eqn.~(\ref{eq:TCR}), but with the lower CR momentum cutoff of
Eqn.~(\ref{eq:qinj}).  In combination with the slope $\alpha_\rmn{inj}$, the
value of $x_\rmn{inj}$ regulates the amount of kinetic energy which is
transferred to the CRs.  Theoretical studies of shock acceleration at galactic
supernova remnants suggest a range of $x_\rmn{inj} \simeq 3.3$ to $3.6$,
implying a particle injection efficiency of $\eta_\rmn{CR,lin} \simeq 10^{-4}$
to $10^{-3}$ \citep{1989A&A...225..179D, 1993ApJ...402..560J,
  1994APh.....2..215B, 1995ApJ...447..944K, 1995A&A...300..605M}.

\begin{figure}
\resizebox{\hsize}{!}{\includegraphics{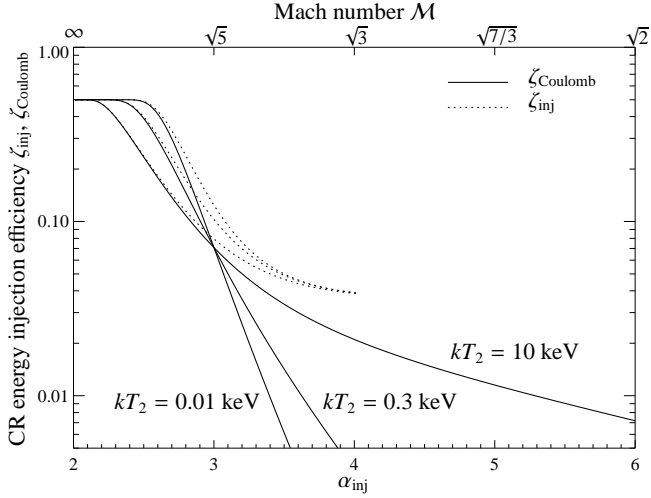}}
\caption{CR energy injection efficiency for the diffusive shock acceleration
  process. Shown is the CR energy injection efficiency $\zeta_\rmn{inj}$
  (dotted) for the three post-shock temperatures $kT_2/\mbox{keV} = 0.01, 0.3$,
  and 10. An effective CR energy efficiency $\zeta_\rmn{Coulomb}$ (solid) is
  obtained by considering Coulomb losses which remove the low-energetic part of
  the injected CR spectrum on a short timescale.  In this figure, we adopted
  the following values for our model parameters, $x_\rmn{inj} = 3.5$,
  $\zeta_\rmn{max} = 0.5$, and $q_\rmn{Coulomb} = 0.03$.  }
\label{fig:fig1}
\end{figure}

Fig.~\ref{fig:fig1} shows the CR energy injection efficiency $\zeta_\rmn{inj}$
as a function of spectral index $\alpha_\rmn{inj}$. It can be clearly seen that
our simplified model for the diffusive shock acceleration fails in the limit of
weak shocks where it over-predicts the injection efficiency. Especially in this
regime, Coulomb and ionisation losses have to be taken into account which
remove the low-energetic part of the injected CR spectrum efficiently on a
short timescale (cf.{\ }Sect.{\ }\ref{sec:Coulomb} \& \ref{sec:Ionisation}).
This gives rise to an effective CR energy efficiency $\zeta_\rmn{Coulomb}$
which is obtained by keeping the normalisation of the CR spectrum fixed while
simultaneously increasing the cutoff: $q_\rmn{inj} \to
q_\rmn{Coulomb}$.\footnote{Solely for illustration purposes, we sketch how
the acceleration efficiency is modified for a constant Coulomb cutoff while we
refer to \citet{Jubelgas} for an algorithm to compute this cutoff on the fly in
simulations.}  Thus, in the linear regime, the effectively injected CR energy
density is given by
\begin{equation}
  \label{eq:zeta_Coulomb}
  \Delta \eps_\rmn{CR,lin,Coulomb} = 
  \Delta n_\rmn{CR,lin} T_\CR(\alpha_\rmn{inj},q_\rmn{Coulomb})
  \left(\frac{q_\rmn{Coulomb}}{q_\rmn{inj}}\right)^{1-\alpha_\rmn{inj}}.
\end{equation}
The effective CR energy efficiency $\zeta_\rmn{Coulomb}$ in the linear regime
is obtained by analogy to the previous considerations,
\begin{equation}
  \zeta_\rmn{lin,Coulomb} =
  \frac{\Delta\eps_\rmn{CR,lin,Coulomb}}{\Delta\eps_\rmn{diss}}.
\end{equation}
Following our suggestion for saturation effects of the shock acceleration given
in Eqn.~(\ref{eq:saturation}), we can obtain the effectively injected CR energy
density in the non-linear regime $\Delta \eps_\rmn{CR,inj,Coulomb}$.  Assuming
a dominant thermal gas component, the spectral index $\alpha_\rmn{inj}$ can
be translated into the Mach number of the shock, $\M_1$, depending on the
adiabatic index of the thermal gas $\gamma$,
\begin{equation}
\M_1 = \sqrt{\frac{2\, (2 + \alpha_\rmn{inj})}
  {1 + 2 \alpha_\rmn{inj} - 3 \gamma}}.
\end{equation}
For a pure thermal gas, the spectral index $\alpha_\rmn{inj} = 2$ formally
corresponds to an infinite Mach number.  In the case of a varying spectral
index, we re-map the changes $\Delta \eps_\rmn{CR,inj}$, $\Delta
n_\rmn{CR,inj}$, and the computed value of the injection spectral index
$\alpha_{\rm inj} = (r + 2)/(r - 1)$ onto the dynamical CR variables
$(C,q,\alpha)$ which describe the total CR population.

\subsubsection{CR injection\label{sec:CRinjStFo} by structure formation shock waves}

For estimating the CR injection at structure formation shock waves in a
numerical simulation a dynamical Mach-number finder is required, which
identifies and characterises shock waves on the fly. Such a Mach number finder
for SPH simulations is presented and tested in \citet{2006MNRAS.367..113P}, and
applied to the problem of CR injection efficiencies of the different shock
waves in the large scale structure.  

\subsubsection{CR injection\label{sec:CRinj} by supernovae}

Shock waves in supernova remnants are believed to be the most important CR
injection mechanism in the galactic context. However, for typical simulations
of galaxy and structure formation the spatial scales of supernovae are below
the resolution length. Therefore we need a sub-grid description for supernova
CR injection.

A significant fraction $\zeta_{\rm SN}\sim 0.1-0.3$ of the kinetic energy of a
supernova may end up in the CR population. Therefore we set the energy
injection rates into the CR and thermal pools to $(\dd\eps_\CR/\dd t)_{\rm SN}
= \zeta_{\rm SN}\, \dd\eps_{\rm SN}/\dd t$ and $(\dd\eps_\th/\dd t)_{\rm SN} =
(1-\zeta_{\rm SN})\, \dd\eps_{\rm SN}/\dd t$, respectively. Here $\dd\eps_{\rm
SN}/\dd t$ is the total SN energy release rate per volume. The increase in CR
number density is given by $(\dd n_\CR/\dd t)_{\rm SN} = (\dd\eps_\CR/\dd
t)_{\rm SN}/T_\CR^{\rm inj}$, where
\begin{equation}
\label{eq:TCR}
T_\CR^{\rm inj} = \mp\,c^2\left[ \frac{q_{\rm inj}^{\alpha_{\rm
inj}-1}}{2}\, \B_\frac{1}{1+q^2} \left(\frac{\alpha-2}{2},
\frac{3-\alpha}{2} \right) + \sqrt{1+q_{\rm inj}^2} -1\right]
\end{equation}
is the average kinetic energy of an injection power-law spectrum with spectral
index $\alpha_{\rm inj}$ and lower momentum cutoff $q_{\rm inj}$. A plausible
value for the injection spectral index is $\alpha_{\rm inj} = 2.4$ in a
galactic context. The low-momentum cutoff can be set to $q_{\rm inj} \sim
\sqrt{kT/(\mp\,c^2)}$ since the power-law spectrum resulting from shock
acceleration reaches down to the thermal population with temperature $kT$.

However, in numerical practice it may be more economical to use a somewhat
higher value for $q_{\rm inj}$, because in many circumstances Coulomb and
ionisation losses will rapidly remove the low energy part of the CR spectrum,
so that the energy of these CRs is nearly instantaneously reappearing as
thermal energy shortly after injection. A slight re-calibration of $\zeta_{\rm
SN}$ can take this into account, so that a numerical code does not have to
explicitly follow the appearance of a short-lived, low energy, super-thermal CR
population.

A criterion to find an adequate $q_{\rm inj}$ is the requirement that the
injected CR spectrum above $q_{\rm inj}$ has a cooling time $\tau_{\rm
cool}(q_{\rm inj}, \alpha_{\rm inj})$ which equals the energy injection
timescale $\tau_{\rm inj}$ defined as the ratio of the present CR energy to the
energy injection rate (above $q_{\rm inj}$).%
\footnote{Another criterion could also be given by the requirement that the
Coulomb-loss timescale $\tau_{\rm C}(q_{\rm inj}) = |T_{\rm p}(q_{\rm
inj})/(\dd T_{\rm p}(q_{\rm inj})/\dd t)_{\rm C}|$ (see
Eqn.~(\ref{eq:coulomb1})) of the particles near the injection cutoff should be
on the order of the simulation time-step. However, this approach has two
potential shortcomings: First, the dynamics becomes formally time-step
dependent, although the Coulomb cooling dynamics will largely compensate for
this. Second, if small numerical timesteps permit $q_{\rm inj} \ll q \sim 1$,
the large number of low energy CRs injected may drag an existing CR population
to lower energies, due to our simplified CR dynamics.}
The rational behind this approach is that injected CRs with $p<q_{\rm inj}$
would not survive energy losses sufficiently long to make a
significant change to the existing CR population given the current injection
rate.  Further mathematical details about this description can be found in
Sec.~\ref{sec:eqform}.

In the case of a varying spectral index, we also have to calculate the pressure
injection rate $\dot{P}_\CR$ for the injection spectrum according to
Eq.~(\ref{eq:Pcr}).


\subsection{CR diffusion\label{sec:diffusion}}

The ubiquitous cosmic magnetic fields prevent charged relativistic particles to
travel macroscopic distances with their intrinsic velocity close to the speed
of light. Instead, the particles gyrate around, and travel slowly along
magnetic field lines. Occasionally, they get scattered on magnetic
irregularities.  On macroscopic scales, the transport can often be described as
a diffusion process if the gyro-radius can be regarded to be small. The
diffusion is highly anisotropic with respect to the direction of the local
magnetic field, characterised by a parallel $\kappa_\|$ and a perpendicular
$\kappa_\perp$ diffusion coefficient. Both are usually functions of location
and particle momentum.

Microscopically, the scattering of the CR on magnetic irregularities of MHD
waves slows down the parallel transport, but allows the perpendicular
transport since it de-places the gyro-centre of the CRs. Therefore, both
microscopic diffusion coefficients depend on the scattering frequency
$\nu_{\rm scatt}$, but with inverse proportionalities: $\kappa_\| \propto
\nu_{\rm scatt}^{-1}$ and $\kappa_\perp \propto \nu_{\rm scatt}$. Particles
are best scattered by MHD waves with a wavelength comparable to the CR's
gyro-radii, which itself depends on the particle momentum. The various
wavelength bands are populated with different strength.  Therefore, the
scattering frequency is usually a function of the particle momentum. The exact
functional dependence is determined by the plasma turbulence spectrum on
scales comparable to the CR gyro-radii.

In this picture \citep[e.g. see][]{1997ApJ...485..655B}, the diffusion
coefficients can be written as
\begin{eqnarray}
\label{eq:diffpar1}
\kappa_\| &=&  \frac{\kappa_{\rm Bohm}}{\eps} \\
\label{eq:kappaperp1}
\kappa_\perp &=& \frac{\eps}{1+ \eps^2} \,\kappa_{\rm Bohm}\,.
\end{eqnarray}
Here, $\eps\ll 1$ is the ratio of the scattering frequency $\nu_{\rm
scatt}$ to the gyro-frequency $\Omega = v/r_g$, and $\kappa_{\rm Bohm}
= v \,r_{g}/3 = v\,p\,\mp\,c^2/(3\,Z\,e\,B)$ is the Bohm diffusion
coefficient.  In most circumstances, $\kappa_\perp$ will be many orders
of magnitude smaller than $\kappa_\|$. Thus, from a microscopic point
of view, CR cross field diffusion seems to be nearly impossible.

Macroscopically, the cross-field particle transport is much faster than the
microscopic diffusion coefficient suggests. The reason for this is that any
small displacement from the initial field line, which a CR achieved by a
perpendicular microscopic diffusion step, can be strongly (often exponentially)
amplified if the CR travels along its new field line \citep{1978PRL...40..38,
  1995A&A...302L..21D}. This is caused by diverging magnetic field lines due to
a random walk in a turbulent environment. This effect should always be present
at some level even if a large-scale mean field dominates the general magnetic
field orientation. The resulting effective diffusion coefficient
$\bar{\kappa}_\perp$ is difficult to estimate from first principles \citep[see
the discussion in][]{2003A&A...399..409E}, but its dependence on the particle
momentum is the same as that of the parallel diffusion coefficient, due to the
dominant role the parallel diffusion plays in the effective cross field
transport. We therefore assume
\begin{equation}
\bar{\kappa}_\perp(p) = \delta_\perp\,\kappa_\|(p)
\end{equation}
with typically $\delta_\perp \sim 10^{-4} - 10^{-2}$
\citep{1999ApJ...520..204G,2003A&A...399..409E}, but see
\citet{2001ApJ...562L.129N} for arguments of a larger $\delta_\perp
\sim 10^{-1}$. In order to be flexible about the underlying MHD
turbulence which fixes the momentum dependency of $\kappa_\|$, we
assume
\begin{equation}\label{eq:kappapar}
\bar{\kappa}_\|(p) = \kappa_\|(p) = \kappa_0 \,
\beta\,p^{d_p}\,\gamma^{d_\gamma} = \kappa_0 \,
p^{d_p+1}\,\gamma^{d_\gamma-1}\,.
\end{equation}
The velocity factor $\beta$ expresses the reduction of diffusion speed for
non-relativistic particles.  For a power-law turbulence spectrum of the form
$E(k)\,\dd k \propto k^{-\alpha_{\rm turb}}\,\dd k$ one obtains $d_p =
2-\alpha_{\rm turb}$ and $d_\gamma = 0$, e.g. $d_p = \frac{1}{3}$ for a
Kolmogorov-type spectrum. We have included the parameter $d_\gamma$ in order
to allow for low-energy deviations from a pure momentum power-law dependence.
Such deviations can for example be caused by modifications of the turbulence
spectrum due to MHD-wave-damping by the low-energy bulk of the CR population
with small gyro-radii. 

It should be noted that turbulent motions in the gas also lead to an
effective diffusion \citep[e.g.][]{2004ApJ...615L..41C}. Due to the
Lagrangian nature of our CR-fluid description, any turbulent diffusion due to
numerically resolved eddies is automatically included. Unresolved turbulence,
however, may have to be taken separately into account by adding a momentum
independent diffusion term to Eqn.~(\ref{eq:kappapar}).

The equation describing  the evolution of the CR spectrum $f(\vecbf{x}, p,
t)$ due to diffusion is 
\begin{equation}
\label{eq:diff_master}
\left(\frac{\partial f}{\partial t}\right)_{\rm diff}  = 
\frac{\partial}{\partial x_i}
\kappa_{ij}\,\frac{\partial f}{\partial x_j}\,,
\end{equation}
where the diffusion tensor
\begin{equation}
\kappa_{ij}(p) = \kappa_\|(p) \,[b_i\,b_j\, +
\delta_\perp\,(\delta_{ij} - b_i\,b_j)] = 
\tilde{\kappa}_{ij} \, p^{d_p+1}\,\gamma^{d_\gamma-1}\,
\end{equation}
is anisotropic with respect to the local main magnetic field direction
$\vecbf{b}(\vecbf{x}) = \vecbf{B}(\vecbf{x})/B(\vecbf{x})$. Since we are
interested in a simplified description, we have to translate
Eqn.~(\ref{eq:diff_master}) into changes of CR number and energy density.
Integrating Eqn.~(\ref{eq:diff_master}) over $p$ leads to an equation governing
the change in CR number density due to diffusion:
\begin{eqnarray}
\label{eq:diff_number}
\left( \frac{\partial n_\CR}{\partial t} \right)_{\rm diff} \!\!\!\! &=&
\frac{\partial}{\partial x_i}
\tilde{\kappa}_{ij}\,\frac{\partial}{\partial x_j} \left\{
\frac{C}{\alpha-2-d_p}\left[ q^{-\alpha+2+d_p}\,\left(
1+q^2\right)^{-\frac{1-d_\gamma}{2}} - \right. \right. \nonumber \\
&& \!\!\!\!\!\! \left. \left. \frac{1-d_\gamma}{2}\,
\B_{\frac{1}{1+q^2}}\left( \frac{\alpha-1-d_p - d_\gamma}{2}, \frac{4+
d_p -\alpha}{2}\right) \right] \right\}.
\end{eqnarray}
This can only lead to reasonable results if the condition $\alpha > 1
+ d_p + d_\gamma$ is fulfilled. For a Kolmogorov-turbulence diffusion
coefficient ($d_p = \frac{1}{3},\, d_\gamma =0$), this translates into
$\alpha>1.33$.

One could assume that the transported energy is simply $(\dd\eps_\CR )_{\rm
  diff} = T_\CR\, (\dd n_\CR )_{\rm diff}$. However, this ansatz would ignore
that the more energetic particles diffuse faster, implying that the effective CR
energy diffusion is more rapid than the CR number diffusion. In order to model
this, we multiply Eqn.~(\ref{eq:diff_master}) by $T_{\rm p}(p)$ and integrate
over $p$. This leads to
\begin{eqnarray}
\label{eq:diff_energy}
\left( \frac{\partial \eps_\CR}{\partial t} \right)_{\rm diff}
\!\!\!\!&=& \frac{\partial}{\partial x_i}
\tilde{\kappa}_{ij}\,\frac{\partial}{\partial x_j} \left\{
\frac{C\,\mp\,c^2}{\alpha-2-d_p} \right. \times\nonumber\\
&& \left. \left[ q^{-\alpha+2+d_p}\,\left( 
\left(1+q^2\right)^{\frac{d_\gamma}{2}} -
\left(1+q^2\right)^{\frac{d_\gamma-1}{2}}
\right) + \right. \right. \nonumber \\
&& \left. \left. \frac{1-d_\gamma}{2}\, \B_{\frac{1}{1+q^2}}\left(
\frac{\alpha-1-d_p - d_\gamma}{2}, \frac{4+ d_p -\alpha}{2}\right)
+ \right. \right. \nonumber \\
&&
 \left. \left.
 \frac{d_\gamma}{2}\, \B_{\frac{1}{1+q^2}}\left(
\frac{\alpha-2-d_p - d_\gamma}{2}, \frac{4+ d_p -\alpha}{2}\right)
\right] \right\}\!.
\end{eqnarray}
This equation can only give reasonable results if the condition $\alpha > 2
+ d_p + d_\gamma$ is fulfilled. For a Kolmogorov-turbulence diffusion
coefficient ($d_p = \frac{1}{3},\, d_\gamma =0$), this translates into
$\alpha>2.33$.

In the case of a varying spectral index, we need to compute $(\partial P_\CR /
\partial t)_\rmn{diff}$ which can be obtained by multiplying
Eqn.~(\ref{eq:diff_master}) by $p\,\beta(p)$ and integrating over $p$. This
results
in
\begin{eqnarray}
\label{eq:diff_P}
\left( \frac{\partial P_\CR}{\partial t} \right)_{\rm diff}
&=& \frac{\partial}{\partial x_i}
\tilde{\kappa}_{ij}\,\frac{\partial}{\partial x_j} 
\frac{C\,\mp\,c^2}{6} \nonumber\\
&\times& \B_{\frac{1}{1+q^2}}\left(
\frac{\alpha-2-d_p - d_\gamma}{2}, \frac{4+ d_p -\alpha}{2}\right).
\end{eqnarray}
This equation can only lead to reasonable results if again the condition
$\alpha > 2 + d_p + d_\gamma$ is fulfilled.

\subsection{CR in-situ re-acceleration\label{sec:insitu}}

The diffusive propagation of CRs implies that CR particles scatter resonantly
on plasma waves with wavelength comparable to their gyro-radii. Since these
waves are propagating, the CRs exchange not only momentum, but also energy with
the waves, leading to a re-acceleration of an existing cosmic ray population.
Since the CR-number is not changed by this process, we can state:
\begin{equation}
\label{eq:reaccncr}
\left(  \frac{\partial n_\CR}{\partial t}\right)_{\rm re-acc} = 0  .
\end{equation}
The change in the CR energy can be derived from the  Fokker-Planck
equation for the 3-dimensional momentum distribution function $f^{(3)}(p) =
f(p)/(4\upi\,p^2)$ of an isotropic CR distribution:
\begin{equation}
\label{eq:re-accf3}
  \left(  \frac{\partial f^{(3)}(p) }{\partial t}\right)_{\rm re-acc} =
  \frac{1}{p^2} \,\frac{\partial }{\partial p} \left( p^2 \, D_p \frac{\partial
  f^{(3)}}{\partial p} \right) \,,
\end{equation}
where $D_p$ is the (pitch-angle averaged) momentum-space diffusion
coefficient. $D_p$ is a function of $p$, which we parametrise by
\begin{equation}
  D_p = D_0 \, p^{1-a_p} \, \gamma^{1-a_\gamma}\,.
\end{equation}
We note that here, as in the section on CR diffusion, a power-law spectrum of
the relevant plasma waves is assumed, as it is expected in the inertial range
of MHD turbulence. In case the CRs are numerous enough to extract energy from
the wave spectrum with a considerable rate (compared to the wave-cascade or
-decay time of the turbulence), the wave spectrum itself would be modified,
reducing the acceleration efficiency. A self-consistent description would
therefore also need to follow the wave spectrum and its modification due to the
energy extraction by CRs \citep[e.g.][]{1996ApJ...461..445M,
2004MNRAS.350.1174B, 2005AdSpR..35..162P, 2007astro.ph..3591B}. However, this
is beyond the scope of this paper. Thus, as soon the cosmic ray population
becomes important due to re-acceleration, our description has
probably left its range of validity.

Taking the appropriate moments of Eqn.~(\ref{eq:re-accf3}) leads to evolution
equations for the CR number density, Eqn.~(\ref{eq:reaccncr}), and for the CR
energy density,
\begin{eqnarray}
  \left(\frac{\partial \eps_\CR}{\partial t}\right)_{\rm re-acc} &=& 
  (2+\alpha)\, C\,D_0\,\mp c^2
  \nonumber\\
  &\times&
  \left[\frac{1}{2}\B_{\frac{1}{1+q^2}}\left(\frac{\alpha+a_p+a_\gamma -2}{2}, 
      \frac{2-\alpha -a_p}{2} \right)\right.
  \nonumber\\
    &+& \left(\sqrt{1+q^2} - 1\right) q^{\alpha-a_p}
    \left(1+q^2\right)^{(1-a_\gamma)/2}\Big]\,,
\end{eqnarray}
which is valid for $\alpha > 2-a_p - a_\gamma$.  This energy has to be taken
from the kinetic-energy dissipation budget of the hydrodynamical flow.
In the case of a varying spectral index, we can obtain the evolution equation
for the CR pressure by taking the appropriate moment of
Eqn.~(\ref{eq:re-accf3}),
\begin{eqnarray}
  \left(\frac{\partial P_\CR}{\partial t}\right)_{\rm re-acc} &=& 
  (2+\alpha)\, C\,D_0\,\frac{\mp c^2}{3}
  \nonumber\\
  &\times&
  \left[\B_{\frac{1}{1+q^2}}\left(\frac{\alpha+a_p+a_\gamma}{2}, 
      \frac{2-\alpha -a_p}{2} \right)\right.
  \nonumber\\
  &+&
  \frac{1}{2}\B_{\frac{1}{1+q^2}}\left(\frac{\alpha+a_p+a_\gamma -2}{2}, 
      \frac{4-\alpha -a_p}{2} \right)
  \nonumber\\
    &+& q^{2-\alpha-a_p}\left(1+q^2\right)^{-a_\gamma/2}\Big]\,,
\end{eqnarray}
which is valid for $\alpha > 2-a_p - a_\gamma$.

The parameterisation of the momentum diffusion coefficient is chosen to be
similar to the one for the spatial diffusion coefficient because of  their
related physical background: both are due to scattering on the same plasma
waves. Quasi-linear calculations of the Fokker-Planck transport coefficient of
charged particles interacting with plasma waves
\citep[see][]{2002cra..book.....S} demonstrate that both, the spatial- and the
momentum-diffusion coefficient depend mainly on the pitch-angle diffusion
coefficient $D_{\mu\mu}= D_{\mu\mu}(p,\mu)$, where $\mu = \cos \theta$ is the
cosine of the pitch-angle $\theta$:
\begin{eqnarray}
  D_p &=& \frac{p^2\,V_{\rm A}^2}{\vel^2}\, \int_{-1}^{1}\!  \dd\mu\,
  D_{\mu\mu}(p,\mu)\\
\kappa_\| &=& \frac{\vel^2}{8} \, \int_{-1}^{1}\! \dd\mu\,
  \frac{(1-\mu^2)^2}{D_{\mu\mu}(p,\mu)}.
\end{eqnarray}
Here, $V_{\rm A}$ is the phase-velocity of the scattering plasma waves which
are usually assumed to be Alfv\'en waves\footnote{Also the stronger damped fast
magneto-sonic waves are discussed as efficient accelerators due to their higher
phase velocity \citep[e.g.][]{1979ApJ...230..373E, 2005MNRAS.357.1313C}.} with
$V_{\rm A} = B/\sqrt{4\upi\,\rho}$. In the relevant inertia ranges of the
MHD-turbulence spectra, it is expected that the amplitude but not the
composition of waves changes with wavelength.\footnote{This wont be the case
anymore as soon some part of the CR population extract a significant amount of
the wave-energy budget from the spectral range they can resonate with. In this
case the CR and wave spectra become interdependent and a decompositon of the
pitch-angle diffusion coefficient into a momentum and a pitch-angle dependent
function is impossible \citep[e.g.][]{1996ApJ...461..445M, 2004MNRAS.350.1174B,
2005AdSpR..35..162P}.} Therefore, the pitch-angle diffusion coefficient should
be separable in $p$ and $\mu$, e.g. $D_{\mu\mu}(p,\mu) = D_1(p) \,
D_2(\mu)$. This allows us to relate the two diffusion coefficients via
\begin{equation}
  D_p \, \kappa_\| = p^2 \,V_{\rm A}^2 X_2\,,
\end{equation}
where $X_2$ is a constant of order unity, and is formally given by
\begin{equation}
  X_2 = \frac{1}{8}\, \left( \int_{-1}^{1}\!  \dd\mu\, D_{2}(\mu) \right)
  \,\left(\int_{-1}^{1}\!  \dd\mu\, \frac{(1-\mu^2)^2}{D_{2}(\mu)} \right)\,.
\end{equation}
Therefore, in the framework of a quasi-linear approximation, the parameters
describing in-situ re-acceleration and diffusion are related by
\begin{equation}
  a_p = d_p\,,\;\; a_\gamma = d_\gamma\,,\;\;\mbox{and}\;\; D_0 = 
  V_{\rm A}^2 X_2/\kappa_0 \,,
\end{equation}
e.g. for a Kolmogorov-like spectrum of Alfv\'en waves $a_p = d_p =
\frac{1}{3},\, a_\gamma = d_\gamma =0$, and $X_2 \sim O(1)$.

\subsection{Coulomb losses}\label{sec:Coulomb}
The energy loss of a proton with $\gamma \ll m_{\rm p}/m_{\rm e}$ by Coulomb
losses in a plasma is given by \cite{1972Physica....58..379G}:
\begin{equation}
\label{eq:coulomb1}
- \left( \frac{\dd T_\p(p)}{\dd t} \right)_{\rm C} = \frac{4 \, \upi
 \, e^4 \,n_{\rm e}}{m_{\rm e}  \, \beta \, c}
 \left[ \ln \left( \frac{2 m_{\rm e} c^2 \beta p}{\hbar
\omega_{\rm pl}} \right) - \frac{\beta^2}{2} \right].
\end{equation}
Here, $\omega_{\rm pl} = \sqrt{4 \upi e^2 n_{\rm e} /m_{\rm e}}$ is the plasma
frequency, and $n_\e$ is the number density of free electrons. We note that
in a neutral gas the Coulomb losses can coarsely be estimated with the
same formulae, provided $n_\e$ is taken to be the total electron number density
(free plus bound). However, atomic charge shielding effects lower the Coulomb
losses significantly. A more accurate description of ionisation losses is given
in Sect. \ref{sec:Ionisation}.

In order to obtain the Coulomb energy losses of the CR population, one has to
integrate Eqn.~(\ref{eq:coulomb1}) over the spectrum $f(p)$. This integration
can certainly be performed numerically. For fast and efficient applications, an
approximative analytical expression might be more practical. We derive such an
expression by replacing the term $\beta\,p$ in the Coulomb logarithm with its
mean value for the given spectrum, which can be written as $\langle \beta\,p
\rangle = 3\,P_\CR/(\mp\,c^2\,n_\CR)$.\footnote{This leads to a slight
  overestimate of the energy losses. A slight underestimate results by setting
  $\langle \beta\,p \rangle \rightarrow q^2/\sqrt{1+q^2}$ in Eqn.
  \ref{eq:coulomb2}. As long these two terms lead to similar loss rates our
  approximately treatment is a good description. Otherwise the integration has
  to be performed numerically.} The Coulomb energy losses are then
\begin{eqnarray}
- \left( \frac{\dd\eps_\CR}{\dd t} \right)_{\rm C} &\approx& \frac{2  \upi C
  e^4 n_{\rm e}}{m_{\rm e} \, c}
 \left[ \ln \left( \frac{2 m_{\rm e} c^2 \,\langle \beta\,p
 \rangle}{\hbar \,\omega_{\rm pl}} \right)\, \B_{\frac{1}{1+q^2}} 
\left( \frac{\alpha-1}{2},- \frac{\alpha}{2}\right) \right. \nonumber\\
&& \left. -\frac{1}{2}\, \B_{\frac{1}{1+q^2}} 
\left( \frac{\alpha-1}{2},- \frac{\alpha-2}{2}\right) 
\right]
\label{eq:coulomb2}
\end{eqnarray}
Since Coulomb losses only affect the lower energy part of the spectrum and
therefore should leave the normalisation unaffected, we propose to set $(\dd
n_\CR/\dd t)_{\rm C} = (\dd\eps_\CR/\dd t)_{\rm C}/T_\p(q)$. This mimics the
effect of Coulomb losses on a spectrum quiet well, since Coulomb losses remove
the lowest energy particles most efficiently from the spectrum, moving them
into the thermal pool. Because also their energy is thermalised, we have
$(\dd\eps_\th/\dd t)_{\rm C} = -(\dd\eps_\CR/\dd t)_{\rm C}$ and $(\dd
n_\p/\dd t)_{\rm C} = - (\dd n_\CR/\dd t)_{\rm C}$. The second equation (which
expresses proton number conservation) can be neglected for convenience due to
the smallness of the effect.  The case of a varying spectral index can be
treated in the same way as for a constant spectral index because Coulomb
losses do not significantly modify the spectral index of the ultra-relativistic
CR population.

The Coulomb loss time scales $\tau = \eps_\CR/(\dot{\eps}_\CR)_{\rm C}$ of CR
populations are shown in \ref{fig:fig_tCool}.

\subsection{Ionisation losses}\label{sec:Ionisation}

Ionisation losses of a proton can be calculated with the Bethe-Bloch equation
\citep{groom-klein2000}, which we cast into the form
\begin{equation}
\label{eq:ionisation1}
- \left( \frac{\dd T_\p(p)}{\dd t} \right)_{\rm I} 
= 
\frac{4 \, \upi \, e^4 }{m_{\rm e} \, \beta \, c} \sum_Z \, Z \,n_{\rm Z} \left[ \ln
 \left( \frac{2 m_{\rm e} c^2 p^2}{I_Z\, b(\gamma)} \right) - \beta^2 -
 \frac{\delta_Z}{2} \right].
\end{equation}
Here, $n_Z$ is the number density of atomic species with electron number $Z$,
and $I_Z$ its ionisation potential. $b(\gamma) = \sqrt{1+ 2\gamma m_{\rm
e}/m_{\rm p} + (m_{\rm e}/m_{\rm p})^2}$ is a minor correction factor for the
here relevant regime $\gamma \ll m_{\rm p}/m_{\rm e}$, which we ignore in the
following. The density correction factor $\delta_Z$ is usually not significant
for gases, but should be given here for completeness:
\begin{equation}
  \delta_Z = \left\{ 
  \begin{array}{ll}
      2\, y - D_Z & y_{1,Z} < y\\
      2\, y - D_Z  + a_Z ((y_{1,Z}-y)/\ln 10)^{k_Z} & y_{0,Z} < y < y_{1,Z}\\
      0                  & y < y_{0,Z}      
  \end{array}
  \right.
\end{equation}
Here, $y = \ln p$, $D_z = 1 - 2 \ln(\hbar \,\omega_{\rm pl}/I_Z)$ and
$y_{0,Z}$, $y_{1,Z}$, $a_Z$, $k_Z$ are empirical constants, which characterise
the atomic species \citep{sternheimer52}. The values for Hydrogen and Helium
can be found in Table \ref{tab:sternheimer}. It is apparent that the density
effect can be neglected below 60~GeV, and therefore we propose to ignore it in
applications in which the CR spectrum is trans-relativistic.

\begin{table}
  \begin{tabular}{lllllll}
\hline
Element & $Z$ & $I_Z$ & $y_{0,Z}$   &  $y_{1,Z}$ & $a_Z$ &  $k_Z$\\
\hline
H$_2$ & 1 & 13.6 eV    & $1.76 \ln 10$& $3 \ln 10$ & 0.34  & 5.01 \\
He & 2 & 24.6 eV       & $2.0 \ln 10$ & $3 \ln 10$ & 0.98  & 4.11 \\
\hline
  \end{tabular}
\caption{\label{tab:sternheimer} Atomic parameters as given by
  \citet{sternheimer52}. The Hydrogen measurements were done with molecular
  hydrogen, but no significant changes for atomic hydrogen are expected with in
  the accuracy required for our description.}
\end{table}

Note that the high energy limit of Eq. (\ref{eq:ionisation1}) is similar, but not
exactly identical to the Coulomb loss formula Eq. (\ref{eq:coulomb1}).

Similar to the case of the Coulomb losses, we insert characteristic
momenta into the logarithmic factors in order to allow for an analytical
integration of the energy losses over the CR spectrum. Since the losses are
dominated by the trans-relativistic regime, we propose to use simply $\langle
p^2 \rangle \approx \langle \beta\,p \rangle = 3\,P_\CR/(\mp\,c^2\,n_\CR)$ and
to set $y = \frac{1}{2} \ln \langle p^2 \rangle$ in order to estimate
$\delta_Z$. This yields
\begin{eqnarray}
- \left( \frac{\dd\eps_\CR}{\dd t} \right)_{\rm I} 
&\approx& 
\frac{2  \upi C
  e^4 }{m_{\rm e} \, c}
\sum_Z \, Z\,n_{\rm Z} 
 \left\{ \left[ \ln \left( \frac{2 m_{\rm e} c^2 \,\langle p^2
 \rangle}{I_z} \right) - \frac{\delta_Z}{2}\right]\, 
\times \right. \\
&& \left. 
\B_{\frac{1}{1+q^2}} 
\left( \frac{\alpha-1}{2},- \frac{\alpha}{2}\right) 
%
- \B_{\frac{1}{1+q^2}} 
\left( \frac{\alpha-1}{2},- \frac{\alpha-2}{2}\right) 
\right\} \nonumber
\label{eq:ionisation2}
\end{eqnarray}

A by-product of the energy losses is the generation of free electrons.  The
production rate of free electrons in a Hydrogen gas can be estimated using the
empirical observation that per $36$ eV lost by the proton on average a free
electron is produced, either as a direct or as a secondary knock-on electron
\citep{1992hea..book.....L}. Therefore, the ionisation rate is
\begin{equation}
  \left(\frac{\dd n_{\rm e}}{\dd t} \right)_{\rm I} \approx -\frac{1}{36\,{\rm eV}} \left(
  \frac{\dd\eps_\CR}{\dd t} \right)_{\rm I} \,.
\end{equation}
The remaining fraction ($1-I_1/36 {\rm eV} = 62\,\%$) of the energy loss is
heating the medium directly. This electron production term can be included in
any ionisation equilibrium calculation for the medium.

\subsection{Bremsstrahlung losses\label{sec:Brems}}
Bremsstrahlung energy losses of protons are usually negligible, since they are
suppressed by a factor $\me^2/\mp^2 \approx 3\, 10^{-7}$ compared with electron
bremsstrahlung losses, which are in turn usually already small.  Therefore, we
do not include bremsstrahlung energy losses of protons in our description.

\begin{figure}
\resizebox{\hsize}{!}{\includegraphics{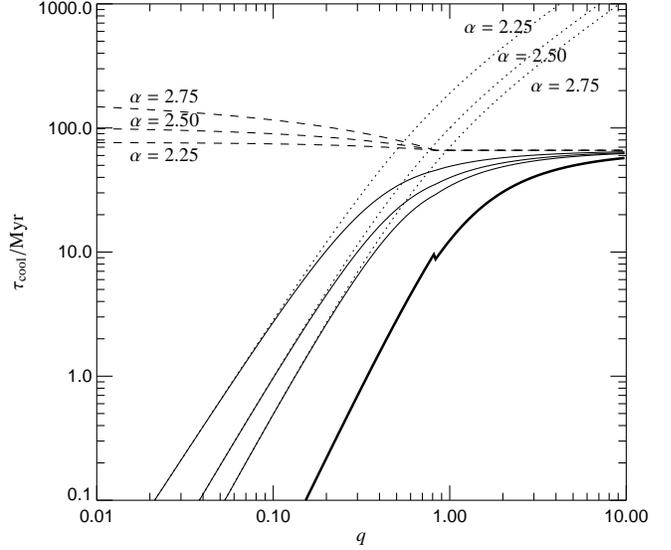}}
\caption{Cooling times of CR spectra with spectral indices $\alpha = $ 2.25,
  2.50, and 2.75 as a function of the cutoff $q$ (thin solid lines) in a pure
  hydrogen plasma with $n_{\rm e} = 1\,{\rm cm^{-3}}$. Coulomb
  and hadronic energy loss times are also displayed with dotted and dashed
  lines, respectively. The thick solid lines gives the total cooling time of a
  mono-energetic CR population with $p=q$. The feature at $q = 0.83$ is due to
  the onset of the hadronic losses above this momentum.}
\label{fig:fig_tCool}
\end{figure}

\subsection{Hadronic losses\label{sec:caloss}}
Another important process is the inelastic reaction of CR nuclei with atoms and
molecules of interstellar and intergalactic matter.  The CR protons interact
hadronically with the ambient thermal gas and produce mainly neutral and
charged pions, provided their momentum exceeds the kinematic threshold
$q_\rmn{thr} m_\p c^2 = 0.78 \mbox{ GeV}$ for the reaction.  The neutral pions
decay after a mean lifetime of $9\times 10^{-17}\mbox{ s}$ into $\gamma$-rays
while the charged pions decay into secondary electrons (and neutrinos):
\begin{eqnarray}
  \pi^\pm &\rightarrow& \mu^\pm + \nu_{\mu}/\bar{\nu}_{\mu} \rightarrow
  e^\pm + \nu_{e}/\bar{\nu}_{e} + \nu_{\mu} + \bar{\nu}_{\mu}\nonumber\\
  \pi^0 &\rightarrow& 2 \gamma \,.\nonumber
\end{eqnarray}
With hadronic interactions, only the CR population above the kinematic
threshold $q_\rmn{thr}$ is visible through its decay products in $\gamma$-rays
and synchrotron emission.  Because of baryon number conservation in strong and
electro-weak interactions, we always end up with pions and two protons in this
CR-proton hadronic interaction (the possibly produced neutron will decay after
a mean lifetime of $886 \mbox{ s}$ into a second proton). Thus, the CR number
density is conserved, implying $(\dd n_\CR/\dd t)_\rmn{had} = 0$.

The total energy loss of CRs is independent of the detailed mechanisms
of how the energy has been imparted on pions during hadronic interactions and
given by
\begin{equation}
  \label{eq:ElossCR}
  -\left(\frac{\dd E_\p}{\dd t}\right)_{\rm had} 
  = c n_\rmn{N} \, \sigma_\rmn{pp} K_\p \, 
  T_\p \, \theta(p_\p - q_\rmn{thr}).
\end{equation}
Here, $\sigma_\rmn{pp}$ is the total pion cross section which is given by
Eqn.~(\ref{sigmapp}), $K_\p \simeq 1/2$ denotes the inelasticity of the
reaction in the limiting regime \citep{1994A&A...286..983M}, and $n_\rmn{N} =
n_\e/(1 - \frac{1}{2} X_\rmn{He})$ denotes the target nucleon density in the
ICM, assuming primordial element composition with $X_\rmn{He} = 0.24$.  The
change in energy density of CRs because of hadronic losses is given by
\begin{eqnarray}
  \label{eq:epslossCR}
  -\left(\frac{\dd \eps_\CR}{\dd t}\right)_{\rm had} &=& 
  \int_0^{\infty}\!\!\!\!\! \dd p\, f(p)\, 
  \left(\frac{\dd E_\p}{\dd t}\right)_{\rm had} \nonumber \\
  &=& c n_\rmn{N}\, \bar{\sigma}_\rmn{pp} K_\p \, 
  \eps_\CR\left(\mbox{max}(q,q_\rmn{thr})\right),
\end{eqnarray}
where $\eps_\CR$ is given by Eqn.~(\ref{eq:epscr}) in which the lower spectral
break $q$ has to be replace by $\mbox{max}(q,q_\rmn{thr})$.  The case of a
varying spectral index can be dealt with in the same way as a constant spectral
index because hadronic losses do not significantly modify the CR spectral
index.  The hadronic loss time scales $\tau = \eps_\CR/(\dot{\eps}_\CR)_{\rm
h}$ of CR populations are shown in \ref{fig:fig_tCool}.

\subsection{Gamma ray emission\label{sec:gammas}}

An analytic formula describing the omnidirectional differential $\gamma$-ray
source function resulting from pion-decay of a power-law CR population is given
in \citet{2004A&A...413...17P}, yielding in our notation:
\begin{eqnarray}
\label{qgamma}
\lefteqn{
s_\gamma(E_\gamma)\,\dd E_\gamma\,\dd V\simeq
\frac{2^4 C}{3 \alpha}\,
\frac{\sigma_\rmn{pp}\,n_\rmn{N}}{m_\p c}\,
\left( \frac{m_\p}{2 m_{\pi^0}}\right)^{\alpha}} \nonumber  \\
  & & \quad\times\,
\left[\left(\frac{2 E_\gamma}{m_{\pi^0}\, c^2}\right)^{\delta} +
      \left(\frac{2 E_\gamma}{m_{\pi^0}\, c^2}\right)^{-\delta}
      \right]^{-\alpha/\delta}\!\dd E_\gamma\,\dd V.
\end{eqnarray}
The formalism also includes the detailed physical processes at the threshold of
pion production like the velocity distribution of CRs, momentum dependent
inelastic CR-proton cross section, and kaon decay channels.  The shape
parameter $\delta$ and the effective cross section $\sigma_\rmn{pp}$ depend on
the spectral index of the $\gamma$-ray spectrum $\alpha$ according to
\begin{eqnarray}
\label{delta}
\delta &\simeq& 0.14 \,\alpha^{-1.6} + 0.44\qquad\mbox{and} \\
\label{sigmapp}
\sigma_\rmn{pp} &\simeq& 32 \cdot
\left(0.96 + \rmn{e}^{4.4 \,-\, 2.4\,\alpha}\right)\mbox{ mbarn}. 
\end{eqnarray}
There is a detailed discussion in \citet{2004A&A...413...17P} how the
$\gamma$-ray spectral index $\alpha_\gamma$ relates to the spectral index of
the parent CR population $\alpha$.  In Dermer's model, the pion multiplicity is
independent of energy yielding the relation $\alpha_\gamma = \alpha$
\citep{1986ApJ...307...47D, 1986A&A...157..223D}.

The derivation of the pion-decay induced $\gamma$-ray source function
implicitly assumed the kinematic threshold $q_\rmn{thr}$ to be above the lower
break of the CR spectrum $q$. This assumption is satisfied in the case of our
Galaxy, where a flattening of the CR spectrum occurs below the kinematic energy
threshold $E_\rmn{thr} = 1.22\mbox{ GeV}$ \citep{1983ARNPS..33..323S}. If the
inequality $q < q_\rmn{thr}$ is violated in the simulation for sufficiently
long timescales, the resulting $\gamma$-ray emission maps have to be treated
with caution.

Provided the CR population has a power-law spectrum, the integrated
$\gamma$-ray source density $\lambda_\gamma$ for pion decay induced
$\gamma$-rays can be obtained by integrating the $\gamma$-ray source function
$s_\gamma(E_\gamma)$ in Eqn.~(\ref{qgamma}) over an energy interval
yielding
\begin{eqnarray}
\label{lambda_gamma}
\lambda_\gamma &=& \lambda_\gamma(E_1, E_2)  = \int_{E_1}^{E_2} \dd E_\gamma\,
s_\gamma(E_\gamma) \\
\label{lambda_gamma2}
&=& \frac{4\, C}{3\, \alpha\delta}
\frac{m_{\pi^0} c\, \sigma_{\rmn{pp}}n_\rmn{N}}{m_\p}
\left( \frac{m_\p}{2 m_{\pi^0}}\right)^{\alpha}\,
\left[\mathcal{B}_x\left(\frac{\alpha + 1}{2\,\delta},
    \frac{\alpha - 1}{2\,\delta}\right)\right]_{x_1}^{x_2},\\
x_i &=& \left[1+\left(\frac{m_{\pi^0}c^2}{2\,E_i}
      \right)^{2\,\delta}\right]^{-1} \mbox{~for~~}
      i \in \{1,2\}.
\end{eqnarray}

\subsection{Hadronically induced synchrotron emission}\label{sec:hadron}
This section describes the hadronically induced radio synchrotron emission
while employing the steady-state approximation for the CR electron spectrum
following \citet{2000A&A...362..151D} and \citet{2004A&A...413...17P}.  This is
only justified if the dynamical and diffusive timescales are long compared to
the synchrotron timescale. This may well be the case in clusters of galaxies,
however, probably not in our own Galaxy.  Possible re-acceleration processes of
CR electrons like continuous in-situ acceleration via resonant pitch angle
scattering by turbulent Alfv\'en waves as well as CR electron injection by
other processes are neglected in this approach.

Assuming that the parent CR proton population is represented by the power-law
of Eqn.~(\ref{eq:spec1}), the CR electron population above a GeV is therefore
described by a power-law spectrum
\begin{eqnarray}
\label{fe_hadr}
f_\e (E_\e) &=& \frac{C_\e}{\rm GeV}
\,\left( \frac{E_\e}{\rm GeV} \right)^{-\alpha_\e}, \\
\label{nCRe}
\mbox{and~~}
C_\e &=& \frac{16^{2-\alpha_\e}}{\alpha_\e - 2}\,
  \frac{\sigma_\rmn{pp}\, m_\e^2\,c^4}
     {\sigma_\rmn{T} \,\rmn{GeV}}
\frac{n_\rmn{N} C}{\eps_B + \eps_\rmn{ph}}
\left(\frac{m_\p c^2}{\mbox{GeV}}\right)^{\alpha-1},
\end{eqnarray}
where the effective CR-proton cross section $\sigma_\rmn{pp}$ is given by
Eqn.~(\ref{sigmapp}), $\sigma_\rmn{T}$ is the Thomson cross section, $\eps_B =
B^2/(8\upi)$ is the local magnetic field energy density, and $\eps_\rmn{ph} =
\eps_\rmn{CMB} + \eps_\rmn{stars}$ is the energy density of the cosmic
microwave background (CMB) and starlight photon field.  $\eps_\rmn{CMB} =
B^2_\rmn{CMB}/(8\upi)$ can be expressed by an equivalent field strength $B_{\rm
CMB} = 3.24\, (1+z)^2\umu\mbox{G}$.  $\eps_\rmn{stars}$ either has to be
guessed or calculated from information of the star distribution, or ignored.

The synchrotron emissivity $j_\nu$ at frequency $\nu$ and per steradian of
such a CR electron population (\ref{fe_hadr}), which is located in an isotropic
distribution of magnetic fields \citep[Eqn.~(6.36) in][]{1979rpa..book.....R},
is obtained after averaging over an isotropic distribution of electron pitch
angles, yielding
\begin{eqnarray}
\label{jnu}
j_\nu&=& A_{E_\rmn{syn}}(\alpha_\e)\, C_\e\,
\left[\frac{\eps_B}{\eps_{B_\rmn{c}}}\right]^{(\alpha_\nu+1)/2}
\propto \eps_\cre\, B^{\alpha_\nu+1} \nu^{-\alpha_\nu},
\\
\label{Bc}
B_\rmn{c} &=& \sqrt{8\upi\, \eps_{B_\rmn{c}}} 
   = \frac{2\upi\, m_\e^3\,c^5\, \nu}{3\,e \mbox{ GeV}^2}
   \simeq 31 \left(\frac{\nu}{\mbox{GHz}} \right)~\umu\rmn{G},
\\
A_{E_\rmn{syn}} &=& \frac{\sqrt{3\upi}}{32 \upi}
   \frac{B_\rmn{c}\, e^3}{m_\e c^2}
   \frac{\alpha_\e + \frac{7}{3}}{\alpha_\e + 1}
   \frac{\Gamma\left( \frac{3\alpha_\e-1}{12}\right)
         \Gamma\left( \frac{3\alpha_\e+7}{12}\right)
         \Gamma\left( \frac{\alpha_\e+5}{4}\right)}
        {\Gamma\left( \frac{\alpha_\e+7}{4}\right)},
\end{eqnarray}
where $\Gamma(a)$ denotes the $\Gamma$-function \citep{1965hmfw.book.....A},
$\alpha_\nu = (\alpha_\e-1)/2 = \alpha/2$, and $B_\rmn{c}$ denotes a (frequency
dependent) characteristic magnetic field strength which implies a
characteristic magnetic energy density $\eps_{B_\rmn{c}}$.

\section{Testing the accuracy of the formalism}\label{sec:tst}

The rigid spectral form of a single power-law with cutoff, which is imposed on
the CR populations in our formalism, leads to inaccuracies compared to the
proper spectral solutions of the evolution equations of the CR spectra. The
question is how large are the inaccuracies for different physical quantities,
and under which circumstances they can or cannot be tolerated.

Two tests are presented: a steady state situation, where cooling balances
injection and a time dependent, freely cooling CR spectrum, without injection.

\subsection{Steady state CR spectrum}\label{sec:steady1}
\subsubsection{Accurate steady state spectrum}
In order to address these questions we analyse a steady state situation in which
accurate and approximated spectra can be worked out analytically. In a
homogeneous environment the CR spectrum $f(p,t)$ should follow the evolution
equation
\begin{equation}\label{eq:spec_evol}
  \frac{\partial f(p,t)}{\partial t} + \frac{\partial}{\partial p} \left(
  \dot{p}(p)\, f(p,t) \right) = Q(p) - \frac{f(p,t)}{\tau_{\rm loss}(p)}\,,
\end{equation}
where 
\begin{equation}
  \dot{p}(p) = \left[ \left( \frac{\dd T_\p(p)}{\dd t} \right)_{\rm C} +
  \left( \frac{\dd T_\p(p)}{\dd t} \right)_{\rm cata} \right] 
  \left(\frac{\dd T_\p(p)}{dp} \right)^{-1}
\end{equation}
is the momentum loss rate due to Coulomb- and hadronic losses\footnote{We
approximate here the hadronic losses as a continuous energy loss.} (see
Eqs.~(\ref{eq:coulomb1})~\&~(\ref{eq:ElossCR})). We assume a fully ionized Hydrogen
plasma with a thermal electron density of $n_{\rm e} = 1 \, {\rm cm}^{-3}$. A
time-independent CR injection spectrum
\begin{equation}
  Q(p) = Q_{\rm inj}\, p^{-\alpha_{\rm inj}}\,,
\end{equation}
with an injection normalisation $Q_{\rm inj} = 1\, {\rm Myr^{-1}}$ (for a here
unspecified volume element) and $\alpha_{\rm inj} = 2.5$ is assumed. The
environment is assumed to be sufficiently extended so that particle escape can
be ignored, implying $\tau_{\rm loss}(p) = \infty$ except for the
thermalisation of CRs at very low momenta.

The steady state spectrum is then given by
\begin{equation}
  f(p) = \frac{Q_{\rm inj}\,p^{1-\alpha_{\rm inj}}}{|\dot{p}(p)|\, (\alpha_{\rm
  inj}-1)} 
\approx \frac{Q_{\rm inj}\,p^{-\alpha_{\rm inj}}}{(\alpha_{\rm
  inj}-1) A_{\rm C}}
\left\{
  \begin{array}{r@{\quad:\quad}l}
    p^{3} & p\ll p_*\\
    p_*^3 & p\gg p_*\\
  \end{array}
\right.\,,
\end{equation}
and its energy distribution can be seen in Fig.~\ref{fig:fig_eqSpec}. The
asymptotic approximations assume negligible hadronic and Coulomb losses in the
low and high energy regimes, respectively. Furthermore, the weak logarithmic
dependence of the Coulomb losses on the particle momentum was ignored in the
asymptotic equations. The cross-over momentum $p_*\approx 1$ depends on the
ratio of the Coulomb to hadronic loss rates coefficients
\begin{eqnarray}
  A_{\rm C} &=&    \frac{4 \, \upi
 \, e^4 \,n_{\rm e}}{m_{\rm e}  \,  c}
\ln \left( \frac{2 m_{\rm e} c^2 \beta p}{\hbar
\omega_{\rm pl}} \right)\\
A_{\rm had} &=& \sigma_\rmn{pp} K_\p \,  m_{\rm p}\,c^3\, n_\rmn{N}
\end{eqnarray}
and is given by
\begin{equation}
  p_* = \sqrt[3]{\frac{A_{\rm C}}{A_{\rm had}}}
= \sqrt[3]{\frac{4 \, \upi \, e^4\, n_\rmn{e}
\ln ( ({2 m_{\rm e} c^2})/({\hbar \omega_{\rm pl}}))
}{
m_{\rm e} \, m_{\rm p}\,c^4 \,\sigma_\rmn{pp} K_\p \, n_\rmn{N}} 
     } \approx 1.087\,.
\end{equation}
Here, we have inserted a fiducial electron density of $n_e = 1\,{\rm cm^{-3}}$
and assumed a pure hydrogen plasma. $p_* \approx 1.1$ is accurate to 10\% for
electron densities in the range $[10^{-6}, 10^{3}]\,{\rm cm^{-3}}$. As a
reference case for an analytic approximation to the steady state equilibrium CR
spectrum we introduce a simple matched asymptotic solution:
\begin{equation}
  f_{\rm approx}(p) \approx \frac{Q_{\rm inj}\,p^{-\alpha_{\rm inj}}}{(\alpha_{\rm
  inj}-1) A_{\rm C}\, (p_*^{-3} + p^{-3})}\,,
\end{equation}
which is also displayed in Fig.~\ref{fig:fig_eqSpec}.

\subsubsection{Approximate steady state spectrum}\label{sec:eqform}

In this section, we calculate the equilibrium spectrum provided by our simplified CR
formalism. Since our formalism is based on CR particle and energy conservation,
the injected particle number and energy should balance the ones due to Coulomb
and hadronic losses.  CR energy balance therefore yields $\dd \eps_\CR/\dd t =
(\dd \eps_\CR/\dd t)_{\rm inj} + (\dd \eps_\CR/\dd t)_{\rm cool}= 0$ which is equivalent to 
\begin{equation}
\label{eq:energy1}
  \left(\frac{\dd \eps_\CR}{\dd
  t} (Q_{\rm inj}, q_{\rm inj}, \alpha_{\rm inj})  \right)_{\rm inj}  =
  \frac{\eps_\CR(C,q,\alpha)}{\tau_{\rm cool} (q,\alpha)} 
\end{equation}
The CR cooling time-scale due to Coulomb and hadronic losses is defined by
\begin{equation}
     \tau_{\rm cool}^{-1}(q, \alpha) = \tau_{\rm C}^{-1}(q, \alpha) + \tau_{\rm had}^{-1}(q,
  \alpha) = 
  \frac{\left(\frac{\dd\eps_\CR}{\dd t} (C, q, \alpha)\right)_{\rm
  C\,+\,had} }{\eps_\CR(C ,q, \alpha)}  .
\end{equation}
The injected number of particles must be balanced by the number of particles
removed from the CR population due to Coulomb losses, since hadronic losses
conserve particle number in the spectrum: $\dd n_\CR/\dd t =
(\dd n_\CR/\dd t)_{\rm inj} + (\dd n_\CR/\dd t)_{\rm C}= 0$. Since $(\dd
n_\CR/\dd t)_{\rm C} = (\dd \eps_\CR/\dd t)_{\rm C}/T_{\rm p}(p)$ we get
\begin{equation}
\label{eq:number1}
  \left(\frac{\dd n_\CR}{\dd
  t} (Q_{\rm inj}, q_{\rm inj}, \alpha_{\rm inj})  \right)_{\rm inj}  =
\frac{\eps_\CR(C,q,\alpha)}{  T_{\rm p}(q) \, \tau_{\rm C}(q,\alpha)}\,.
\end{equation}
Dividing Eq.~(\ref{eq:energy1}) by Eq.~(\ref{eq:number1}) yields
\begin{equation}
\label{eq:qqi}
  T_\CR(q_{\rm inj}, \alpha_{\rm inj})  =
  T_{\rm p}(q) \left( 1+ \frac{\tau_{\rm C}(q,\alpha)}{\tau_{\rm
  had}(q,\alpha)} \right)\,,
\end{equation}
which is free of any normalisation constant ($C$, $C_{\rm inj}$) and defines an
implicit function relating the injection cutoff $q_{\rm inj}$ and the
equilibrium cutoff $q=q(q_{\rm inj}, \alpha_{\rm inj}, \alpha)$.

\begin{figure*}
\resizebox{\hsize}{!}{\includegraphics{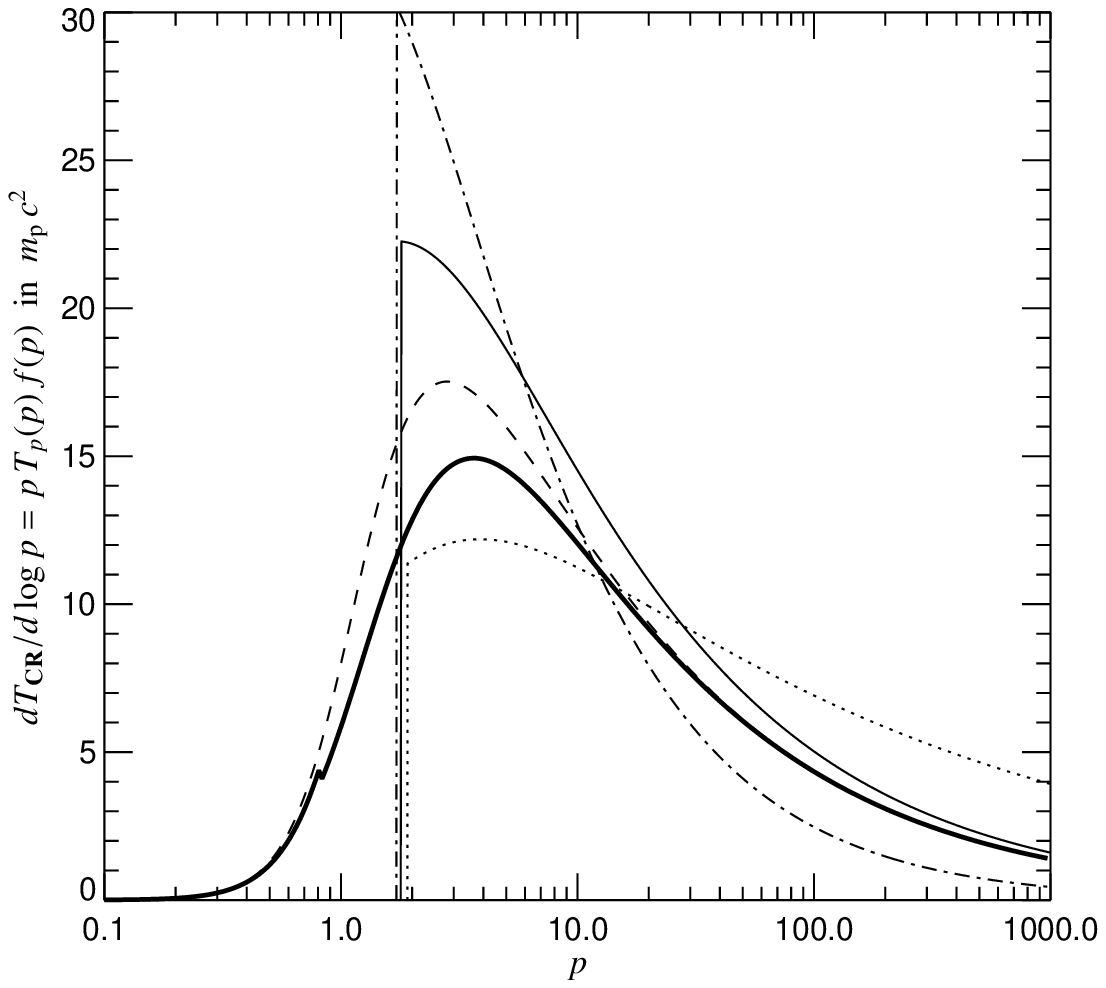}
\includegraphics{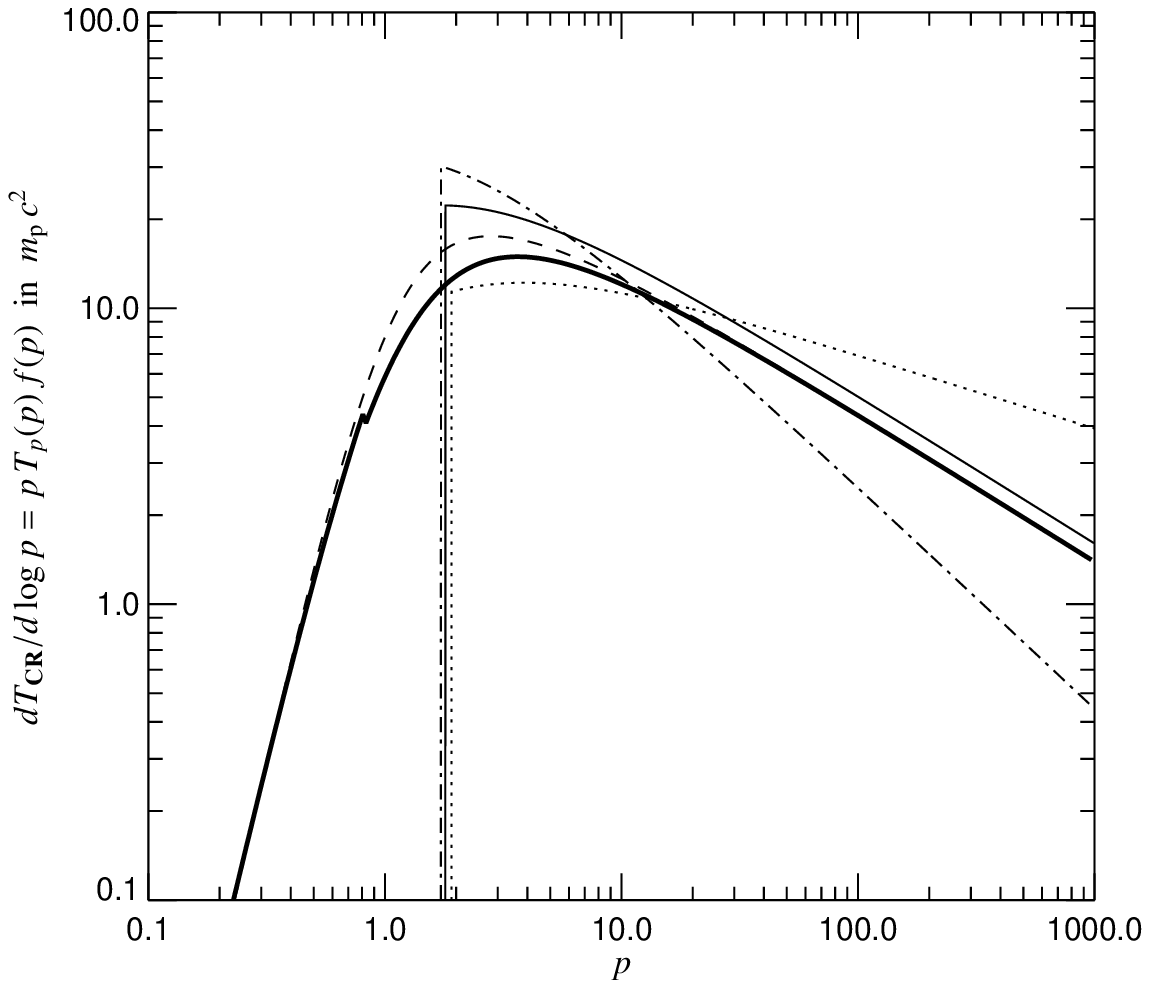}}
\caption{Spectral energy distribution per logarithmic momentum interval of a
  steady state CR population within a volume element of our test system in
  log-linear (left) and log-log (right) representation. CRs
  are injected with an spectral index of $\alpha_{\rm inj} = 2.5$. The thick
  line gives the correct spectrum of the full problem, the dashed line is the
  matched asymptotic solution, and the profiles with the low-momentum cutoff
  are equilibrium spectra in our formalism for the assumed spectral indices
  $\alpha = 2.25$ (dotted curve), $\alpha = \alpha_{\rm inj} = 2.5$ (solid
  line), and $\alpha = 2.75$ (dash-dotted curve). The small bump at $p=0.83$ in
  the correct spectrum is due to vanishing hadronic losses below this momentum.
  The areas under the curves in this plot are proportional to the CR energy,
  which differ less than 20\% between approximate and accurate spectra (see
  Tab.~\ref{tab:accuracy}).}
\label{fig:fig_eqSpec}
\end{figure*}

In order to specify $q$ fully a second condition between $q$ and $q_{\rm inj}$ is
required. A very simplistic, but -- as we will see a posteriori -- sometimes
reasonable choice, is the condition $q_{\rm inj} = q$.  However, our
criterion to specify $q_{\rm inj}$ is the requirement that the injected CR
spectrum above $q_{\rm inj}$ has a cooling time $\tau_{\rm cool}(q_{\rm inj},
\alpha_{\rm inj})$ which equals the energy injection time-scale $\tau_{\rm
inj}$ defined as the ratio of the present CR energy to the energy injection
rate (above $q_{\rm inj}$).  Therefore, we have
\begin{eqnarray}
  \tau_{\rm cool} &=& \tau_{\rm inj}\,,{\rm where}\\
  \tau_{\rm cool} &=&  \tau_{\rm cool}(q_{\rm inj}, \alpha_{\rm inj}) = 
  \frac{\eps_\CR(\dot{C}_{\rm inj} ,q_{\rm inj}, \alpha_{\rm inj})}{(\dd\eps_\CR(\dot{C}_{\rm
  inj} ,q_{\rm inj}, \alpha_{\rm inj})/\dd t)_{\rm C\,+\,had}}\,, \,\mbox{and}
  \nonumber \\
\tau_{\rm inj} &=& \tau_{\rm inj}(q_{\rm inj}, \alpha_{\rm inj}, C, q, \alpha) =
\frac{\eps_\CR(C ,q, \alpha)}{(\dd\eps_\CR( \dot{C}_{\rm
  inj} ,q_{\rm inj}, \alpha_{\rm inj})/\dd t)_{\rm inj}}\,.\nonumber
\end{eqnarray}
This leads to the condition
\begin{equation}
\label{eq:time1}
  \tau_{\rm cool}(q_{\rm inj}, \alpha_{\rm inj}) = \tau_{\rm cool}(q, \alpha)\,, 
\end{equation}
which indeed implies $q_{\rm inj} = q$ in the case $\alpha_{\rm inj} = \alpha$. In
other cases one has to solve Eqs.~(\ref{eq:qqi}) and (\ref{eq:time1}) numerically
for $q_{\rm inj}$ and $q$.  The normalisation of the equilibrium CR spectrum can
then be derived from Eq.~(\ref{eq:number1}) and is given by
\begin{equation}
  C = Q_{\rm inj}\, \tau_{\rm C}(q,\alpha)\, \frac{T_{\rm p}(q)}{T_\CR(q,\alpha)}\,
  \frac{1 -\alpha}{1 -\alpha_{\rm inj}} \frac{q_{\rm inj}^{1 -\alpha_{\rm
  inj}}}{q^{1 -\alpha}} \,.
\end{equation}
The resulting spectra for $\alpha_{\rm inj} = 2.5$ and $\alpha = 2.25$, $2.5$,
and $2.75$ are displayed in Fig.~\ref{fig:fig_eqSpec}.

\subsubsection{Comparison of the spectra}\label{sec:steady3}

Although Fig.~\ref{fig:fig_eqSpec} shows that the approximated and correct CR
spectra are different in detail, their integral properties are very
similar. The total energy in the approximated spectrum with $\alpha =
\alpha_{\rm inj} = 2.5$ is only 10\% above the correct one, and the pressure
difference is even smaller with 5\%. Since the low energy part of the spectrum
is ignored in our formalism, it is not surprising that the total CR number
density is underestimated by 60\%. Similarly, the Coulomb heating of the
thermal background gas by the approximated CR spectrum seems to be
underestimated by 80\%. However, this is not correct, since in our formalism
the CRs injected below the cutoff $q$ return their energy instantaneously to
the thermal gas. If one takes this into account one finds that the Coulomb
heating rates agree with 1\% accuracy. Since our description is energy
conserving, this implies that the hadronic losses, and thereby the total
hadronic induced radiation in gamma rays, electrons and neutrinos, agree also
on a 1\% level. However, the agreement of the differential radiation spectra at
a given photon, electron or neutrino energy is worse ($\sim 10\% \ldots 30\%$
overestimate), due to the different normalisation of the high energy part of
the approximated and the correct CR spectra.

For comparison, we also calculate the accuracy of the matched asymptotic
spectrum and equilibrium spectra of our formalism for spectral indices
different from the injection index, namely $\alpha = 2.25$ and $\alpha =
2.75$. The corresponding relative accuracies of CR energy, pressure, number,
and Coulomb loss rates are similar and can be found in Tab.~\ref{tab:accuracy}.

This demonstrates that, despite the roughness of the representation of the CR
spectrum in our formalism, CR energy density and pressure are calculated with
an accuracy acceptable for first exploration studies of the influence of CRs on
galaxy and structure formation. The limitations of the approach, especially in
representing the detailed spectral shape of the CR population, have also become
clearer.

\begin{table}
\include{table}
\caption{Relative accuracy of CR energy, pressure, number, and Coulomb loss
  rate in the various spectral approximations displayed in
  Fig.~\ref{fig:fig_eqSpec}, compared to the numerical solution of the problem
  (also displayed in Fig.~\ref{fig:fig_eqSpec}).}
\label{tab:accuracy}
\end{table}

\begin{figure*}
\resizebox{\hsize}{!}{\includegraphics{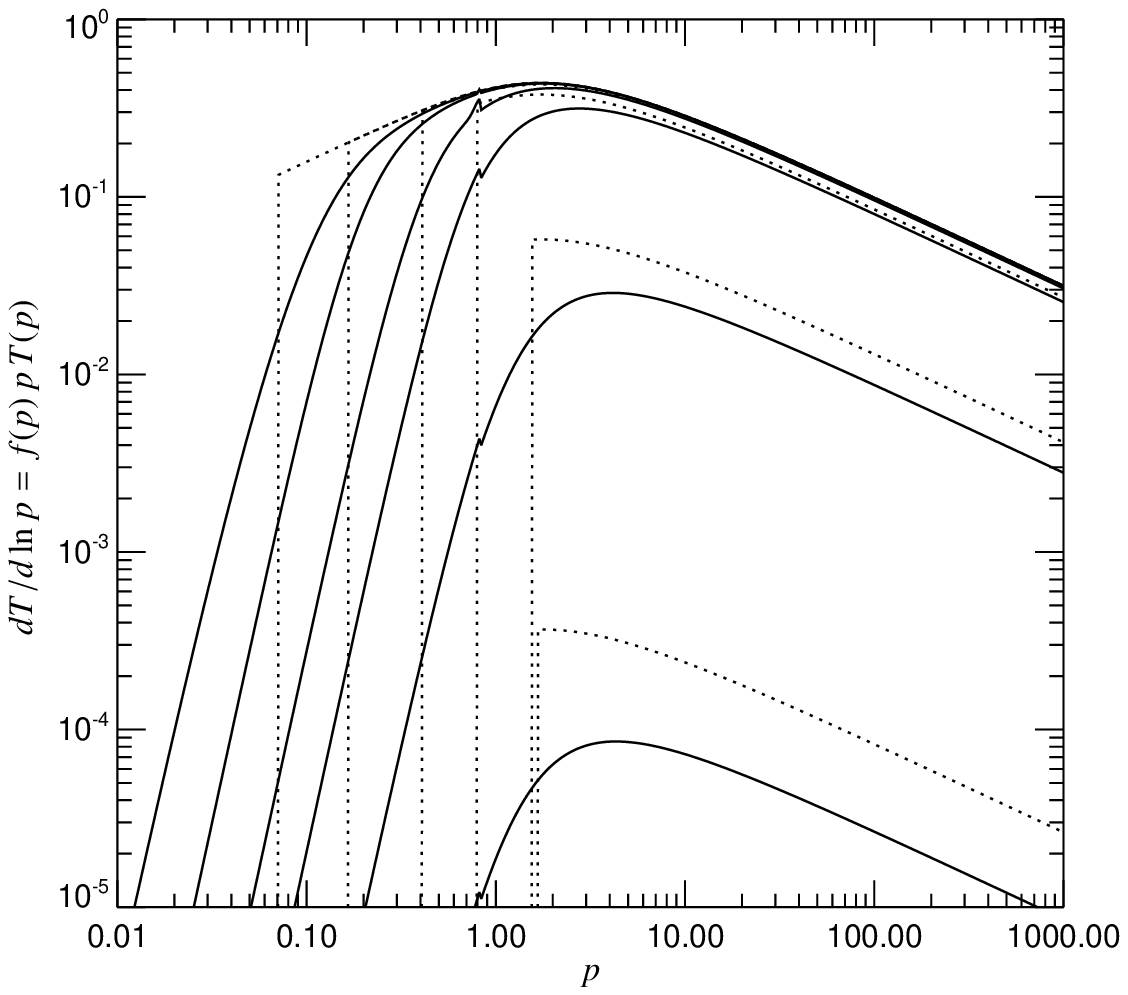}\includegraphics{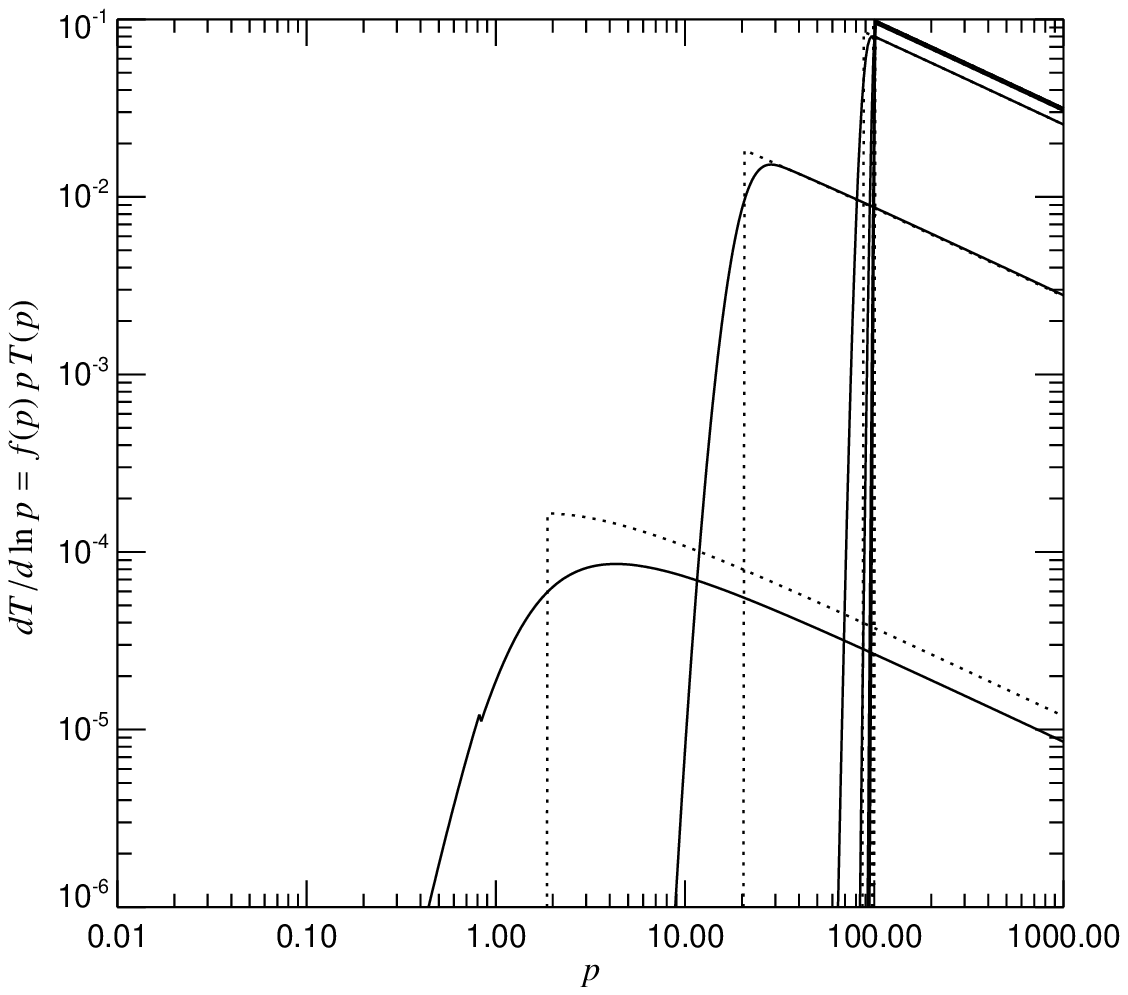}}
\caption{Spectral energy distribution per logarithmic momentum interval of a
  freely cooling CR populations within a volume element of our test system. The initial CR
  populations are described by $(C_0,q_0, \alpha) = (1, 10^{-3}, 2.5)$ (left panel) and 
  $(C_0,q_0,\alpha) = (1, 10^{3}, 2.5)$ (right panel). Spectra are shown for cooling ages of 
  $\approx (0.01, 0.1, 1, 10, 100, 300)$ Myr. The numerically exact solutions are shown with solid 
  lines. The approximative solutions are displayed by dotted lines.}
\label{fig:fig_coolSpec}
\end{figure*}

\subsection{Freely cooling CR spectra}

 Setting the right-hand side in Eqn. (\ref{eq:spec_evol}) to zero
describes the cooling of an initially injected
spectrum. Fig. \ref{fig:fig_coolSpec} shows the time evolution of
numerically exact and approximate spectra with an initial condition
described by $(C_0,q_0, \alpha) = (1, 10^{-3}, 2.5)$ and
$(C_0,q_0,\alpha) = (1, 10^{3}, 2.5)$, respectively. The numerically
exact spectra were obtained using the CR spectrum evolution code by
Jasche et al.~(2007, submitted). The approximative CR spectra were
calculated using the implementation of the approximative CR evolution
by \citet{Jubelgas}. A fully ionized Hydrogen-only plasma with
electron density of $1\,{\rm cm^{-3}}$ was assumed.

For timescales below the cooling time of $\sim 50$ Myr most of the spectral
evolution happens in the low-energy cutoff, and the approximate treatment
captures the evolution of the exact solution quite well. Both solution approach
a similar asymptotic shape, with a low energy cut off close to the
trans-relativistic regime. However, the cooling rate seems to be different for
the two descriptions, the approximate solution seems to cool on a slightly
larger timescale. This inaccuracy is a consequence of the approximative
description. 

The momentum losses per hadronic interaction are lower for
trans-relativistic particles than for ultra-relativistic particles. In
our approximative description the ultra-relativistic particles benefit
therefore from the reduced energy losses of the trans-relativistic
CRs. This effect should therefore vanish for an
ultra-relativistic-only CR spectrum. Indeed, as can be seen in the
right panel of Fig. \ref{fig:fig_coolSpec}, as long as all CRs are
ultra-relativistic, the approximative solution follows very closely the
exact one.  

To summarize, within one hadronic cooling time, the approximative spectral
evolution is accurate if the correct spectral index is set. However, the
long-term evolution of a freely cooling CR population becomes inaccurate after
a few cooling times. This is the more the case the steeper the CR spectrum
is. For applications where first order effects of the CR dynamics are
considered, this level of accuracy should be sufficient. For applications which
require a high level of accuracy over several cooling times a proper treatment
of the spectral evolution would be needed. However, this is beyond the scope of
this work.


\section{Smoothed particle hydrodynamics}\label{sec:sph}

In this section, we describe how the dynamical effects of a CR
population can be included into the smoothed particle hydrodynamics
(SPH) simulation technique.

\subsection{Lagrangian fluid dynamics}\label{eq:lagrange}

The Lagrange density of a magneto-hydrodynamical gas-CR medium is
\begin{equation}
{\mathcal L}(\vecbf{r}, \dot{\vecbf{r}}) = 
\frac{1}{2}\, \rho\, \dot{\vecbf{r}}^2 - \eps_\th(\rho,A) -
\eps_\CR(\rho, C_0, q_0) - \eps_B\,,
\end{equation}
where $\eps_\th = \rho\,A_0\,(\rho/\rho_0)^{\gamma-1}/(\gamma-1)$ is the
thermal energy density of a gas with adiabatic index $\gamma$ and an entropy
described by the adiabatic invariant $A$. Any adiabatic invariant $X\in \{A_0,
C_0, q_0\}$ is simply advected with an adiabatic flow:
\begin{equation}
\frac{\dd X}{\dd t} = \left( \frac{\dd X}{\dd t} \right)_{\rm non-adiabatic}\,,
\end{equation}
where the rhs allows for non-adiabatic changes discussed in
Sect.~\ref{sec:nadiab}.  The density evolves according to
\begin{equation}
\dd \ln \rho /\dd t = - \mathbf{\nabla} \cdot \dot{\vecbf{r}}\,,
\end{equation}
and the magnetic field according to the MHD-induction law:
\begin{equation}
\frac{\partial \vecbf{B}}{\partial t} = \mathbf{\nabla} \times (
\dot{\vecbf{r}} \times \vecbf{B} - \eta\,\mathbf{\nabla} \times \vecbf{B})
\end{equation}

\subsection{SPH formulation}\label{sec:sphform}

In smoothed particle hydrodynamics (SPH), the fluid is discretised as
an ensemble of particles which carry the mass, energy, and
thermodynamic properties of the fluid elements. Macroscopic properties
of the medium such as the density at position $\vecbf{r}_i$ of the
$i$-th particle are calculated with adaptive kernel estimation in the
form
\begin{equation}
\rho_i = \sum_j m_j\, W(|\vecbf{r}_i - \vecbf{r}_j|, h_i)\,,
\end{equation}
where $m_j$ is the mass of the $j$-th fluid element and $W(r,h)$ is
the SPH smoothing kernel. The SPH particle positions $\vecbf{r}_i$ are
the dynamical variables of the simulation.  However, following the
approach of \citet[][]{2002MNRAS.333..649S} which we extend in this
work to include an additional CR population and to allow for a general
equation of state of the gas, the smoothing-kernel lengths $h_i$ will
be considered as dynamical variables of a Lagrangian function.

We introduce the CR spectrum of the $i$-th SPH particle
\begin{equation}
m_i\,\hat{f}_i (p) = m_i\, \frac{\dd N_{\CR}}{\dd p\,\dd m} =
\frac{m_i}{\rho(\vecbf{r}_i)} \,\frac{\dd N_{\CR}}{\dd p\,\dd V} =
\frac{m_i}{\rho(\vecbf{r}_i)} \, f_i(p)
\end{equation}
with the help of the CR number per momentum and unit gas mass
$\hat{f}_i(p)$. Our power-law template spectra are then described by
\begin{equation}
  \hat{f}_i(p) = \hat{C}_i \, p^{-\alpha}\, \theta(p- q_i),
\end{equation}
where $\hat{C}_i = C(\vecbf{r}_i) /\rho(\vecbf{r}_i)$ denotes the CR
normalisation constant of the $i$-th SPH particle.  Similarly, we introduce the
CR energy, CR density, and thermal energy per unit gas mass with
$\hat{\eps}_\CR = \eps_\CR/\rho$, $\hat{n}_\CR = {n}_\CR/\rho$, and
$\hat{\eps}_\th ={\eps}_\th/\rho$, respectively. The equations defining these
quantities and their changes due to adiabatic and non-adiabatic processes in
terms of $\hat{C}$ and $q$ can be obtained from the corresponding formulae in
this paper by replacing $C$ with $\hat{C}$. For instance, Eqn.~(\ref{eq:epscr})
yields
\begin{eqnarray}
\label{eq:epscrhat}
\hat{\eps}_{\CR,i} &=& \int_0^\infty \!\!\!\! \!\! \dd p\, \hat{f}_i(p) \,T_{\rm
p}(p)=\frac{\hat{C}_i\, \mp\,c^2}{\alpha-1} \, \times
\nonumber \\
&& \left[\frac{1}{2}
\, \B_{\frac{1}{1+q_i^2}} \left(
\frac{\alpha-2}{2},\frac{3-\alpha}{2}\right) + q_i^{1-\alpha}
\left(\sqrt{1+q_i^2}-1 \right) \right] \,,
\end{eqnarray}
and so on.

The Lagrange formalism provides an elegant way for deriving the
equations of motions for SPH simulations.  The SPH discretised
Lagrangian is
\begin{equation}
{\mathcal L}(\vecbf{q},\dot{\vecbf{q}}) = \sum_i \frac{m_i}{2} \,
\dot{r}_i^2 - \sum_i m_i\, \hat{\eps}_i + \sum_i \lambda_i \,
\phi_i\,,
\end{equation}
where $\hat{\eps}_i = \hat{\eps}_\th + \hat{\eps}_\CR$ is the total
energy per mass of the $i$-th SPH particle, and $\vecbf{q} =
(\{\vecbf{r}_i\}, \{h_i\}, \{\lambda_i\})$, which should not to be
confused with the CR spectral cutoff $q$, denotes all degrees of
freedom of the system. These are the components of the SPH particle
positions $\vecbf{r}_i$ and the smoothing lengths $h_i$ and their
velocities. The $\lambda_i$s are Lagrange multipliers introduced by
\citet[][]{2002MNRAS.333..649S} in order to incorporate the choice of
the smoothing length $h_i$ into the Lagrangian via the function
\begin{equation}
\label{eq:M_SPH}
\phi_i(\vecbf{q}) = \frac{4\,\upi}{3} \,h_i^3\,\rho_i - M_{\rm SPH}\,,
\end{equation}
where $M_{\rm SPH}$ is the required mass within the smoothing kernel. 

Here, we have ignored the description of magnetic fields within the
SPH-Lagrangian. For the moment, we treat the evolution of the magnetic
field separately from this Lagrangian formalism, adding instead the
magnetic forces ad-hoc to the momentum equations of the SPH
particles. This is along the lines of \citet{1999A&A...348..351D}, and
seems to work well in typical cosmological settings. However, we note
that the dynamical influence of magnetic fields can in principle also
be included into the SPH Lagrange-function as
\citet{2004MNRAS.348..123P,2004MNRAS.348..139P} demonstrate.

If one derives the SPH-equation of motion from a Lagrangian, one
obtains a dynamical system which obeys energy and entropy
conservation. Non-adiabatic processes, like shock waves, radiative
energy losses and energy exchange of the thermal and relativistic
fluids, thermal conduction, and CR diffusion have to be added into
these equations. The way such process should be implemented in case of
CR populations should become clear from this work.

\subsection{Equations of motion}

The equations of motions (of the adiabatic, non-magnetic part) of the
SPH description follow from the Hamilton principle, namely
\begin{equation}
  \frac{\dd}{\dd t} \frac{\partial \mathcal{L}}{\partial \dot{\vecbf{q}}} -
  \frac{\partial \mathcal{L}}{\partial \vecbf{q}} = 0\,.
\end{equation}


The equation determining the smoothing length of the $i$-th particle
follows from the variation of the action with respect to the
Lagrange-multiplier $\lambda_i$. The corresponding part of the
Euler-Lagrange equations yields
\begin{equation}
\phi_i = 0\,,
\end{equation}
which for the special form chosen in Eqn.~(\ref{eq:M_SPH}) leads to an
implicit equation for $h_i$ that has to be solved numerically in
practice.

Variation with respect to $h_i$ leads to an equation for $\lambda_i$:
\begin{equation}
\lambda_i = \frac{\partial \hat{\eps}_i}{\partial h_i} \left[
\frac{\partial \phi_i}{\partial h_i}\right]^{-1} = \frac{\partial
  \hat{\eps}_i}{\partial \rho_i} \frac{\partial \rho_i}{\partial h_i} \left[
\frac{\partial \phi_i}{\partial h_i}\right]^{-1}.
\end{equation}
Using now Eqn.~(\ref{eq:M_SPH}), one gets
\begin{equation}
\lambda_i = \frac{3\,m_i}{4\,\upi\,h_i^3} \frac{\partial
  \hat{\eps}_i}{\partial \rho_i}\, g_i\,,\;\;\mbox{with}\;\; g_i \equiv
  \left[1+\frac{h_i}{3\rho_i}\frac{\partial\rho_i}{\partial h_i}
  \right]^{-1}.
\end{equation}
Furthermore, the thermodynamical pressure is defined as
\begin{equation}
  P = - \left(\frac{\partial E}{\partial V} \right)_S\,,
\end{equation}
where $S=S_i$ denotes the entropy of a SPH particle volume element of size
$V=V_i = \rho_i/m_i$ and internal energy $E= m_i\,\hat{\eps}_i$. This pressure
definition can be used to express the derivative of the SPH particle energy
with respect to the local density in terms of the total (thermal plus CR)
thermodynamical pressure:
\begin{equation}
\label{eq:epsrhoP}
  \frac{\partial \hat{\eps}_i}{\partial \rho_i} = \frac{P_i}{\rho_i^2}
  = \frac{P_{th,i}+P_{\CR,i}}{\rho_i^2}
\end{equation}
One might argue that this derivation should only be correct for
thermodynamic systems, and therefore not necessarily for CR
populations which do not exhibit a Boltzmann distribution
function. However, the concept of entropy, and the concept of
adiabatic processes, which do not change phase space density and
therefore leave entropy constant, is well defined for an arbitrary
distribution function. Therefore, Eqn.~(\ref{eq:epsrhoP}) is a
generally valid result, which can also be confirmed by an explicit
calculation.\footnote{The internal energy per SPH particle of an ideal
(thermal and/or relativistic) gas can be written as $m_i\,
\hat{\eps}_i = m_i\, \sum_{a} \int \dd p \, \hat{f}_{a,i}(p)\,T_a(p)$,
where $a$ is the index over the particle species (electrons, protons,
etc.)  with momentum-space distribution functions $\hat{f}_{a,i}(p)$,
and $T_a(p)$ the relativistic correct kinetic energy of the particles
(Eqn.~(\ref{eq:Tcr})). A straightforward calculation of ${\partial
\hat{\eps}_i}/{\partial \rho_i}$, which uses the first equality in
Eqn.~(\ref{eq:Pcr}), leads then to Eqn.~(\ref{eq:epsrhoP}).}

Thus, the Lagrange-formalism for a variable SPH smoothing length
introduced by \citet[][]{2002MNRAS.333..649S} for a polytropic
equation of state can easily be generalised to a general equation of
state by replacing the thermal gas pressure by the total pressure of
all fluid components. A calculation along the lines of
\citet[][]{2002MNRAS.333..649S} shows that the SPH-particle equations
of motion read
\begin{equation}
  \frac{\dd\bvel_i}{\dd t} = - \sum_{j} m_j \left[
  g_i\,\frac{P_i}{\rho_i^2} \mathbf{\nabla}_i\, W_{ij}(h_i) +
  g_j\,\frac{P_j}{\rho_j^2} \mathbf{\nabla}_i\, W_{ij}(h_j) \right]\,,
\end{equation}
with $\bvel_i = \dot{\vecbf{r}}_i$, $P_i= P_{th,i}+P_{\CR,i}$ the
thermal plus CR pressure, and $W_{ij}(h_i) = W(|\vecbf{r}_i -
\vecbf{r}_j|, h_i)$.

\section{Conclusion}\label{seq:concl}

We have introduced a simplified model for the description of cosmic ray physics
with the goal of facilitating cosmological hydrodynamical simulations that
self-consistently account for the dynamical effects of CRs during structure
formation.  We have shown how various adiabatic and non-adiabatic processes can
be described in terms of a simple model for the local CR spectrum, consisting
of a power-law with varying normalisation and low-energy cutoff. In our basic
model, the CR spectral index is held fixed and chosen in advance to resemble a
typical spectral index for the system under consideration. We have also
described an extended model where the spectral index is allowed to vary as
well, which leads to a numerically more involved scheme.

We have demonstrated that dynamical quantities like CR energy, pressure, and
energy loss rates are all reasonable well represented by our approximative treatment
of the CR spectra in case of a steady state situation in which CR injection and
cooling balance each other. The accuracy is $\sim 10\%$ even if the spectral
indices of CR injection and population do not match. Given the large
uncertainties in our knowledge of the parameters determining the CR physics
like CR injection efficiency, the level of accuracy of our numerical treatment
seems to be sufficient. Thus, the formalism is suitable for explorations of the
possible dynamical impact of CRs on galaxy and large-scale structure formation.

We have explained how our description of the CR physics can be
self-consistently included into the SPH simulation methodology. The
formulation we propose manifestly conserves energy and particle
number. In particular, CR entropy is exactly conserved in adiabatic
processes, and the dynamical forces from CR pressure gradients are
derived from a variation principle.

This paper has outlined a basic framework for future work on the
impact of CR populations on galaxy and large-scale structure
formation. It is accompanied by two papers describing the
implementation and testing of (i) the CR formalism as described here
in the GADGET simulation code \citep{Jubelgas}, and (ii) a shock
capturing method for SPH that allows on-the-fly estimation of the Mach
number of structure formation shock waves, which is essential to
follow CR injection accurately \citep{2006MNRAS.367..113P}. Further
science applications are in preparation.

\appendix

\section{Numerically updating the CR spectrum}\label{sec:CRupdate}

\subsection{Constant spectral index}
\label{sec:updateconst}

Updating the adiabatic invariant variables $\hat{C}_0$ and $q_0$ is most
conveniently done via Eqn.~(\ref{eq:dC0}) and (\ref{eq:dq0}). However, if the
relative changes during a numerical time-step are large, e.g. due to rapid CR
production at a location without a substantial initial CR population, these
equations would have to be integrated on a refined time-grid, or solved with an
implicit integration scheme. Both methods would be very time-consuming.
Therefore, we propose another updating scheme: from the initial variables
$\hat{C}_0(t_0)$ and $q_0(t_0)$ at time $t_0$, the corresponding
instantaneously particle number $\hat{n}_\CR$, energy density $\hat{\eps}_\CR$,
and average particle energy $T_\CR$ are calculated according to
Eqns.~(\ref{eq:adiabatic}) to (\ref{eq:Tcr}). Then $\hat{n}_\CR$,
$\hat{\eps}_\CR$, and $T_\CR$ are updated using the non-adiabatic CR energy and
number losses or gains during that time-step. And finally, these updated values
have to be translated back into updated values of $\hat{C}_0(t_1)$ and
$q_0(t_1)$. This is easiest by first inverting Eqn.~(\ref{eq:Tcr}) in order to
calculate $q$, and then to use Eqns.~(\ref{eq:adiabatic}) and (\ref{eq:ncr}) to
get the updated $\hat{C}_0(t_1)$ and $q_0(t_1)$. The inversion of
Eqn.~(\ref{eq:Tcr}) has to be done numerically for $T_\CR \sim \mp\,c^2$, e.g.
using pre-calculated numerical tables. However, for the asymptotic regimes we
propose the following approximate inversion formulae:
\begin{eqnarray}
  q(\tau) &=& 
  \left\{ \begin{array}{ll}
      q_a +  \frac{\dps q_a^{4-\alpha}}{\dps (3-\alpha) \, \mathcal{B}},
      & \tau = T_\CR /(\mp\,c^2) \ll 1 \\
      \rule{0cm}{0.6cm}
      \frac{\dps \alpha -2}{\dps \alpha -1} \,(\tau+1),
      & \tau = T_\CR /(\mp\,c^2) \gg 1 \\
    \end{array} \right. ,\\
  &&\mbox{with}\quad 
  q_a =
  \left(\frac{2\,\tau}{\mathcal{B}}\right)^\frac{1}{\alpha-1},
  \mbox{and}\,\, \mathcal{B}=\mathcal{B}\left(\frac{\alpha-2}{2},
    \frac{3-\alpha}{2}\right).
\end{eqnarray}

\subsection{Variable spectral index}
\label{sec:updatevariable}

If the relative changes of the dynamic CR variables $\hat{C}$, $q$, and
$\alpha$ during a numerical time-step are large we propose the following
numerically efficient updating scheme (instead of updating via
Eqns.~(\ref{eq:variablealpha}) and (\ref{eq:invJacobian})): from the initial
variables $\hat{C}_0(t_0)$, $q_0(t_0)$, and $\alpha(t_0)$ at time $t_0$, the
corresponding instantaneous particle number $\hat{n}_\CR$, energy density
$\hat{\eps}_\CR$, and pressure $\hat{P}_\CR$ are calculated according to
Eqns.~(\ref{eq:adiabatic}) to (\ref{eq:Pcr}). Then $\hat{n}_\CR$,
$\hat{\eps}_\CR$, and $\hat{P}_\CR$ are updated according to the losses or
gains of non-adiabatic CR energy, pressure, and number density during that
time-step. And finally, these updated values have to be translated back into
updated values of $\hat{C}_0(t_1)$, $q_0(t_1)$, and $\alpha(t_1)$. This can be
done by numerically solving the following equation for the root $q$,
\begin{eqnarray}
  \label{eq:updatevar}
  \frac{P_\CR}{n_\CR}\,\,\frac{q^{1-\alpha(q)}}{\alpha(q) - 1} &=& 
  \frac{m_\p c^2}{6}\B_{\frac{1}{1+q^2}}
  \left(\frac{\alpha(q) - 2}{2},\frac{3 - \alpha(q)}{2}\right),
   \mbox{ and}\\
  \label{eq:updatealpha}
  \alpha(q) - 1 &=& \frac{3 P_\CR}{\eps_\CR - T_\p(q) n_\CR}.
\end{eqnarray}
These equations are obtained by combining Eqns.~(\ref{eq:ncr}),
(\ref{eq:epscr}), and (\ref{eq:Pcr}). The new CR spectral index $\alpha$ and
$C$ are obtained by Eqns.~(\ref{eq:updatealpha}) and (\ref{eq:ncr}).

\section*{Acknowledgements}
We acknowledge helpful comments by Eugene Churazov and by an anonymous
referee. We thank Jens Jasche for performing simulations of freely cooling CR
spectra.



\label{lastpage}

\end{document}